\documentclass[aps,prd,twocolumn,superscriptaddress,nofootinbib]{revtex4}
\usepackage{hyperref}
\usepackage{amsmath,amssymb,amsfonts}
\usepackage{color}
\usepackage{graphicx}
\usepackage{enumerate} 
\usepackage{colordvi} 
\usepackage{bm}
\usepackage{multirow}
\usepackage{braket}
\usepackage{dcolumn}

\newcommand{\dlnPk}{n}

\newcommand{\nn}{\nonumber}
\newcommand{\ipmu}{Kavli Institute for the Physics and Mathematics of the Universe (WPI), UTIAS, The University of Tokyo, 5-1-5 Kashiwanoha, Kashiwa, Chiba 277-8583, Japan}

\begin{document}

\title{Information content in anisotropic cosmological fields: Impact of different multipole expansion scheme for galaxy density and ellipticity  correlations}

\author{Takuya Inoue}
\email{tinoue@asiaa.sinica.edu.tw}
\affiliation{Department of Physics, National Taiwan University, No. 1, Section 4, Roosevelt Road, Taipei 106216, Taiwan}
\affiliation{Academia Sinica Institute of Astronomy and Astrophysics (ASIAA), No. 1, Section 4, Roosevelt Road, Taipei 106216, Taiwan}
\author{Teppei Okumura}
\affiliation{Academia Sinica Institute of Astronomy and Astrophysics (ASIAA), No. 1, Section 4, Roosevelt Road, Taipei 106216, Taiwan}
\affiliation{\ipmu}
\author{Shohei Saga}
\affiliation{Institute for Advanced Research, Nagoya University, Furo-cho Chikusa-ku, Nagoya 464-8601, Japan}
\affiliation{Kobayashi-Maskawa Institute for the Origin of Particles and the Universe,
Nagoya University, Chikusa-ku, Nagoya, 464-8602, Japan}
\author{Atsushi Taruya}
\affiliation{\ipmu}
\affiliation{Center for Gravitational Physics and Quantum Information, Yukawa Institute for Theoretical Physics, Kyoto University, Kyoto 606-8502, Japan}

\date{\today}
\begin{abstract}
Multipole expansions have been often used for extracting cosmological information from anisotropic quantities in observation.
However, which basis of the expansion is best suited to quantify the anisotropies is not a trivial question in any summary statistics. In this paper, using the Fisher matrix formalism, we investigate the information content in multipole moments of the power spectra of galaxy density and intrinsic ellipticity fields in linear theory from the Alcock-Paczynski effect and redshift-space distortions (RSD). We consider two expansion schemes, the associated Legendre basis as well as the standard Legendre basis conventionally used in literature. 
We find that the information in the multipoles of the intrinsic alignment (IA) power spectra in the associated Legendre basis converges more slowly to that in the full 2D power spectra than in the Legendre basis.
This trend is particularly significant when we consider a high number density sample.
In this high number density case, we show that the errors on the Hubble parameter obtained from the multipoles of the IA cross- and auto-power spectra in the associated Legendre basis are respectively about $6 \%$ and $10 \%$ larger than the full 2D case even when we use multipoles up to $\ell = 6$.
Our results demonstrate that the choice of basis is arbitrary but changes the information content encoded in multipoles depending on the sample and statistics under consideration.
\end{abstract}

\maketitle

\section{Introduction}
In galaxy redshift surveys, probing the nature of dark energy or testing modifications of gravity is a key scientific goal to explain the current accelerated expansion of the Universe in cosmology. 
Measurements of baryon acoustic oscillations (BAO) \cite{1970ApJ...162..815P,1998ApJ...496..605E,1999MNRAS.304..851M,2003ApJ...598..720S} and redshift-space distortions (RSD) \cite{1977ApJ...212L...3S,1980lssu.book.....P,1987MNRAS.227....1K,1998ASSL..231..185H,2008APh....29..336L} in galaxy redshift surveys have simultaneously constrained the history of cosmic expansion and structure formation \cite{2011PhRvD..83j3527T,2014MNRAS.439.2515O,2012MNRAS.426.2719R}. In doing so, the two-point statistics of the tracer fields containing meaningful cosmological information, such as galaxy density fields, have played a major role as one of the representative summary statistics.

It has been recently advocated that the intrinsic alignments (IAs) of galaxy shapes, which have been long considered as a contamination of weak lensing analysis \cite{2000ApJ...545..561C,2001MNRAS.320L...7C,2000MNRAS.319..649H,2004PhRvD..70f3526H} (see Refs.~\cite{Troxel:2015,Joachimi:2015,Kiessling:2015,Kirk:2015,Lamman:2024} for reviews), can be used as an additional cosmological probe \cite{2012PhRvD..86h3513S,2015JCAP...10..032S,2016PhRvD..94l3507C,2022MNRAS.515.4464C}. 
It has been shown by Refs.~\cite{2020ApJ...891L..42T,2022PhRvD.106d3523O} that combining the galaxy clustering with the IA improves dynamical and geometric constraints. 
Larger contributions from the IA are expected in upcoming surveys with the higher number density and higher quality galaxy shapes.

The IA statistically traces tidal fields of the large-scale matter distribution.
The simplest model to describe the statistics of IA is the linear alignment (LA) model \cite{2001MNRAS.320L...7C,2004PhRvD..70f3526H, 2011JCAP...05..010B, 2015JCAP...08..015B,2020MNRAS.494..694O, 2020MNRAS.493L.124O}, which linearly relates the ellipticity fields observed by each galaxy image with matter density fields. 
Note that similar to galaxy clustering in redshift space, the statistics of IA exhibit anisotropies along the line-of-sight direction, but the IA appears distorted even without RSD, providing an alternative route to improve cosmological constraints.
To maximize the cosmological information in the anisotropic nature of the IA statistics, Refs.~\cite{2020ApJ...891L..42T,2022PhRvD.106d3523O} used the LA model for the power spectra in full 2D characterized by the wave number $k$ and the directional cosine $\mu$ between the line-of-sight direction and wavevector, namely $P(k,\mu)$, 
for the forecast studies.

As known in studies of the galaxy clustering, since the actual analysis in full 2D is complicated by a large amount of data (e.g., Ref.~\cite{2008ApJ...676..889O}), 
the multipole expansion, $P_{\ell}(k)$ \cite{1992ApJ...385L...5H,1994MNRAS.267..785C}, has been frequently used instead of the full 2D spectrum. 
As shown in Ref.~\cite{2011PhRvD..83j3527T}, measuring multipole moments up to the hexadecapole moment (i.e., $\ell=4$) provides the information comparable to the full 2D power spectrum. Their finding was based on the galaxy density auto-power spectrum, but it is not obvious how well multipoles of the IA power spectra have the information relative to the full 2D spectra.

In order to obtain the multipoles of the IA correlations, the Legendre polynomials have been conventionally used as a basis to expand the IA correlations \cite{2016MNRAS.457.2301S,2020MNRAS.493L.124O,2022MNRAS.515.4464C,2022MNRAS.516..787Z}. However, Ref.~\cite{2022PhRvD.105l3501K} revealed that the associated Legendre polynomials provides a more natural basis for IA to characterize the anisotropies originated from the projection of the galaxy shapes onto the sky (see also Refs.~\cite{2023MNRAS.518.4976S,2023PhRvD.108h3533K,2024MNRAS.528.1487S,2024MNRAS.530.3515S}). Furthermore, Ref.~\cite{2024PhRvD.109j3501O} has recently derived new analytic formulas of multipoles in terms of the associated Legendre basis for both linear and nonlinear IA correlations.
In terms of the information content for the cosmological statistical analysis, the choice of the optimal basis for the multipole expansions of anisotropic fields can be a general issue, but has not been well discussed.

Focusing on the specific case, IA power spectra, in this paper, we quantitatively investigate the information content obtained from multipoles in the different bases, especially through the effects of geometric distortions and RSD. 
Considering the three-dimensional galaxy density and ellipticity fields, we provide linear theory expressions of the IA power spectra and analytic expressions for the multipole covariances in the associated Legendre basis, which are needed for the forecasts.
Ref.~\cite{2016MNRAS.457.1577G} mentioned that even if the power spectrum multipoles vanish above a certain multipole, the higher-order multipole covariances do not vanish. Therefore, we show the impact of the higher-order multipole contributions in the covariances on cosmological constraints. Moreover, using multipoles instead of the full 2D spectrum, we simultaneously obtain the geometric and dynamical constraints combining the galaxy clustering with IA and then show the convergence of multipole contributions to the full 2D case.

The rest of this paper is organized as follows. 
In Sec.~\ref{sec:multipoles},  we introduce the multipoles for the IA power spectra in the linear theory limit and present the analytic formulas of the multipole covariance in the associated Legendre basis. In Sec.~\ref{sec: Forecast_formalism}, we describe the forecast formalisms used for our numerical computations. We investigate the impact of the basis on the signal-to-noise ratios and the parameter constraints in Sec.~\ref{sec: results} with further discussions in Sec.~\ref{sec: discussion}. Our conclusions are given in Sec.~\ref{conclusion: conclusion}.
In Appendix~\ref{app: mathematical relations}, we present useful relations between the IA power spectrum and its multipoles in the associated Legendre basis used in Sec.~\ref{sec: detectability}.
In Appendix~\ref{app: derivation cov}, we provide the multipole covariances for the IA power spectra in terms of the associated Legendre basis. In Appendix~\ref{app: dev_power_spectra}, we provide the derivatives of the power spectra and multipoles with respect to parameters.

In this paper, we assume a flat $\Lambda {\rm CDM}$ model, in which our fiducial values of the cosmological parameters are taken from the 7 year Wilkinson Microwave Anisotropy Probe results \cite{2011ApJS..192...18K}.

\section{Multipole moments in terms of the associated Legendre basis in linear theory}
\label{sec:multipoles}
Ref.~
In this section, we present the formalism to deal with IA statistics in the linear theory limit. In Sec.~\ref{sec:multipole power spectrum}, adopting the linear alignment (LA) model
which linearly relates the ellipticity field with the tidal field of large-scale structure \citep{2001MNRAS.320L...7C,2004PhRvD..70f3526H}, 
we provide analytic formulas of the multipole moments of the galaxy density-ellipticity cross-power spectrum and ellipticity-ellpticity auto-power spectrum in redshift space expanded in terms of the associated Legendre basis
These expressions are also summarized in Appendix C of Ref.~\cite{2024PhRvD.109j3501O}. In Sec.~\ref{sec:APeffect}, we consider the geometric distortions by Alcock-Paczynski effect, which  arises from the apparent mismatch of the underlying cosmology when we convert the redshift and angular position of each galaxy to the comoving radial and transverse distances. 
We then provide analytic expressions of their Gaussian covariance matrices in Sec.~\ref{sec:covariance}. 

Throughout this paper, we works with the plane-parallel approximation in which the line-of-sight direction is fixed to a specific direction. 
We omit the redshift dependence in all the equations.

\subsection{Power spectrum multipoles}
\label{sec:multipole power spectrum}

In galaxy imaging surveys, we observe ellipticities of galaxies projected onto the sky. The ellipticity field is characterized by two independent degrees of freedom ($\gamma_{+},\gamma_{\times}$) and measured from the traceless part of the second moments of the observed intensity. The observed ellipticity field $\gamma_{(+/ \times)}$ receives two contributions, the intrinsic ellitpicity $\gamma^{\rm I}_{(+/ \times)}$ and gravitational shear induced by the lensing effect $\gamma^{\rm G}_{(+/ \times)}$. In this paper, we ignore the latter because it is subdominant when taking correlations in a specific redshift slice \cite{2022PhRvD.106d3523O}. 

In the LA model, the ellipticity field, which is primarily given as a traceless rank-2 tensor sourced by the scalar gravitational potential, is characterized by the two-component fields in Fourier space as \cite{2020MNRAS.493L.124O, 2021MNRAS.501..833K} 
\begin{equation}
\gamma_{(+, \times)}(\bm{k}) =  b_{\rm K}\Bigl(1-\mu^2\Bigr)
\Bigl(\cos{2\phi},\sin{2\phi}\Bigr)\delta_{\rm m}(\bm{k})
\label{eq: def gamma_Fourier},
\end{equation}
where $\delta_{\rm m}(\bm{k})$ is the linear matter density field, which is related to the gravitational potential via the Poisson equation, $\mu$ is the directional cosine between the wavevector and the line of sight direction, and $\phi$ is the azimuthal angle of the wavevector projected on the celestial sphere. The quantity $b_{\rm K}$ characterizes the strength of galaxy alignment, and is regarded as  a shape bias parameter. Conventionally, the parameter $A_{\rm IA}$ has been also used in the literature. These are related to each other through
\cite{2015JCAP...10..032S, 2021MNRAS.501..833K,2022PhRvD.106d3523O,2023MNRAS.518.4976S},  
\begin{equation}
b_{\rm K} = -0.0134 A_{\rm IA}\Omega_{\rm m,0}/D_{+}(a),
\label{eq: AIA to bK}
\end{equation}
with $D_+$ and $a$ being, respectively, the linear growth factor and the scale factor of the universe. In Ref.~\cite{2021MNRAS.501..833K}, the redshift dependence of the amplitude $A_{\rm IA}$ was found fairly weak for dark matter halos. This is irrespective of whether halos are selected for a certain mass range or a fixed number density.

Under the plane-parallel approximation, the ellipticity field is naturally decomposed into E/B-modes to be rotational invariance through
\begin{equation}
\gamma_{\rm E}(\bm{k})+i\gamma_{\rm B}(\bm{k})=e^{-2i\phi}\Bigl\{\gamma_{+}(\bm{k})+i\gamma_{\times}(\bm{k})\Bigr\}. \label{eq:EBmodes}
\end{equation}
Substituting Eq.~(\ref{eq: def gamma_Fourier}) into 
Eq.~(\ref{eq:EBmodes}) leads to $\gamma_{\rm E}(\bm{k})=b_{\rm K}(1-\mu^2)\delta_{\rm m}(\bm{k})$ and $\gamma_{\rm B}(\bm{k})=0$. 
As long as we consider the linear theory limit, the power spectra associated with the $\rm B$-mode vanish \cite{2015JCAP...08..015B, 2021MNRAS.501..833K}.
To extract the cosmological information from the galaxy density and ellipticity fields, we calculate the auto- and cross-power spectra of these fields. In the linear theory limit, the galaxy density field $\delta_{\rm g}$ in redshift space is given as \cite{1987MNRAS.227....1K,1992ApJ...385L...5H,1994MNRAS.267..785C},
$
\delta_{{\rm g}}(\bm{k})=\left(b+f\mu^2\right)\delta_{\rm m}(\bm{k})$,
where $b$ and $f$ are, respectively, the linear galaxy bias and the linear growth rate defined by $f\equiv d{\rm ln} D_+(a)/d{\rm ln}a$.
Note that unlike the galaxy density field, the observed ellipticity field is not affected by RSD at linear order, and this is the reason why it does not depend on the growth rate in Eq.~(\ref{eq: def gamma_Fourier})
\cite{2016MNRAS.457.2301S, 2020MNRAS.493L.124O}.

Given the galaxy density and ellipticity fields, we have the auto-power spectra of the density and $\rm E$-mode and their cross-power spectrum, defined by
\begin{align}
\Braket{\delta_{\rm g}(\bm{k})\delta_{\rm g}(\bm{k}')} &= (2\pi)^{3}\delta_{\rm D}(\bm{k}+\bm{k}')P^{\rm g g}(\bm{k}), \label{eq:def GG P}
\\
\Braket{\gamma_{\rm E}(\bm{k})\gamma_{\rm E}(\bm{k}')} &= (2\pi)^{3}\delta_{\rm D}(\bm{k}+\bm{k}')P^{\rm E E}(\bm{k}), \label{eq:def II P}
\\
\Braket{\delta_{\rm g}(\bm{k})\gamma_{\rm E}(\bm{k}')} &= (2\pi)^{3}\delta_{\rm D}(\bm{k}+\bm{k}')P^{\rm g E}(\bm{k}). \label{eq:def GI P}
\end{align}
The resulting expressions are then given as \cite{2020ApJ...891L..42T, 2021MNRAS.501..833K, 2022PhRvD.106d3523O}, 
\begin{align}
P^{\rm gg}(k,\mu)
&=(b+f \mu^2)^2P_{\rm L}(k), \label{eq:GG P}
\\
P^{\rm EE}(k,\mu)&=b^2_{{\rm K}}(1-\mu^2)^2P_{\rm L}(k), \label{eq:II P}
\\
P^{\rm gE}(k,\mu)
&=b_{{\rm K}}(b+f \mu^2)(1-\mu^2)P_{\rm L}(k), \label{eq:GI P}
\end{align}
where $P_{\rm L}$ is the linear matter power spectrum defined by  $\Braket{\delta_{\rm m}(\bm{k})\delta_{\rm m}(\bm{k}')} = (2\pi)^{3}\delta_{\rm D}(\bm{k}+\bm{k}')P_{\rm L}(k)$.
As seen in Eqs.~(\ref{eq:GI P}) and ~(\ref{eq:II P}), they are, respectively, proportional to $(1-\mu^2)$ and $(1-\mu^2)^2$ due to the projection of the ellipticity field along the line of sight.

To characterize anisotropies in the power spectra given above, the multipole expansion via the Legendre polynomials, denoted by $\mathcal{L}_{\ell}(\mu)$, has been frequently used in the literature. However, for the IA power spectra involving the projection factors, the expansion in terms of the associated Legendre polynomials $\mathcal{L}^{m}_{\ell}(\mu)$ with $m>0$ is convenient and they could provide a more natural basis. The IA power spectra are thus expressed as 
\begin{align}
P^{\rm XY}(k,\mu)=\sum_{\ell \geq m} \widetilde{P}^{\rm XY}_{\ell,m}(k)\Theta_{\ell,m}(\mu), \label{eq:multipole expansion}
\end{align}
where $\rm XY= \{gE,EE\} $ and
$\Theta_{\ell, m}(\mu)$ is the normalized associate Legendre polynomials as a basis function in the multipole expansions \cite{Matsubara_2024,2024PhRvD.109j3501O}, 
defined by 
\begin{align}
\Theta_{\ell, m}(\mu)\equiv\sqrt{2 \pi}Y_{\ell, m}(\theta, 0)=
\sqrt{\frac{2 \ell+1}{2}\frac{(\ell-m)!}{(\ell+m)!}}\mathcal{L}^m_{\ell}(\mu).
\label{eq: Normalized associate Legendre}
\end{align}
It satisfies $\Theta_{\ell, -m}(\mu)=(-1)^m\Theta_{\ell, m}(\mu)$ and $\Theta_{\ell, m}(-\mu)=(-1)^{\ell+m}\Theta_{\ell, m}(\mu)$, as well as a simple orthonormal relation, 
$
\int^{1}_{-1}{d}\mu
\Theta_{\ell,m}(\mu)
\Theta_{\ell',m}(\mu)=\delta^{\rm K}_{\ell,\ell'} 
$,
with $\delta^{\rm K}_{\ell,\ell'}$ being the Kronecker delta.
Eq.~(\ref{eq:multipole expansion}) holds for the arbitrary choice of $m$. 
Here, the quantities with tilde represent those expanded in terms of the normalized associated Legendre polynomials, not the 
(conventional) Legendre polynomials. Making use of the orthonormality, the coefficients are obtained by  
\begin{align}
\widetilde{P}^{\rm XY}_{\ell,m}(k) &=\int^{1}_{-1}{d}\mu \Theta_{\ell,m}(\mu) P^{\rm XY}(k,\mu)
. \label{eq: plm def}
\end{align}
In this paper, we call these coefficients the multipoles expanded in terms of the associated Legendre polynomials. Accordingly, those with $m=0$ are referred to as multipoles in terms of the conventional Legendre polynomials, though $\mathcal{L}_{\ell}(\mu)=\mathcal{L}_{\ell}^{0}(\mu) = \sqrt{2/(2\ell+1)}\Theta_{\ell,0}(\mu)$.
Some useful relations hold for the coefficients, as shown in Appendix \ref{app: mathematical relations}.

Substituting Eq.~(\ref{eq:GI P}) into Eq.~(\ref{eq: plm def}), the multipole moments of the gE power spectrum are concisely expressed as 
\begin{equation}
\widetilde{P}^{\rm g E}_{\ell,m}(k) = A^{\rm gE}_{\ell,m}b_{\rm K}P_{\rm L}(k)
. \label{eq:PgE_lm} 
\end{equation}
When choosing the Legendre basis, $m=0$, the non-vanishing coefficients $A^{\rm gE}_{\ell,0}$ are given by
\begin{align}
&
\left(A^{\rm gE}_{0,0},A^{\rm gE}_{2,0},A^{\rm gE}_{4,0}\right)
\nn \\  &=\left(\frac{2 \sqrt{2}}{15}(5b+f), 
\frac{2}{21}\sqrt{\frac{2}{5}}(f-7b), 
-\frac{8 \sqrt{2}}{105}f\right)
. \label{eq:multipole PgE0}
\end{align}
If we choose the associated Legendre basis with $m=2$, we have two non-zero multipoles and the coefficients $A^{\rm gE}_{\ell,2}$ ($\ell=2,4$) are given by
\begin{align}
&
\left(A^{\rm gE}_{2,2},A^{\rm gE}_{4,2}\right)=\left(\frac{4}{7 \sqrt{15}}(7b+f), \frac{8}{21 \sqrt{5}}f \right)
. \label{eq:multipole PgE2}
\end{align}

Similarly to the gE power spectrum, substituting Eq.~(\ref{eq:II P}) into Eq.~(\ref{eq: plm def})
leads to the multipole moments of the EE power spectrum:
\begin{equation}
\widetilde{P}^{\rm E E}_{\ell,m}(k) = A^{\rm EE}_{\ell,m}b^2_{\rm K}P_{\rm L}(k)
~ \label{eq:PEE_lm} 
\end{equation}
with the non-vanishing moments up to $m=4$ summarized as follows:
\begin{align}
\left(A^{\rm EE}_{0,0},A^{\rm EE}_{2,0},A^{\rm EE}_{4,0}\right)&=\left(\frac{8 \sqrt{2}}{15}, -\frac{16}{21}\sqrt{\frac{2}{5}}, \frac{8 \sqrt{2}}{105}\right)
, \label{eq:multipole PEE0}
\\
\left(A^{\rm EE}_{2,2},A^{\rm EE}_{4,2}\right)
&=\left(\frac{8}{7}\sqrt{\frac{3}{5}}, -\frac{8}{21 \sqrt{5}}\right), \label{eq:multipole PEE2} \\
A^{\rm EE}_{4,4}&=\frac{16}{3 \sqrt{35}}
. \label{eq:multipole P}
\end{align}

As increasing $m$, the linear-order contributions become compressed into a fewer numbers of multipoles in the associated Legendre basis. For the EE power spectrum, we have only a single multipole for $m=4$. Therefore, in the following analysis, we will treat $m=4$ as the representative associated Legendre basis for the EE power spectrum. 
It is, however, not obvious whether the choice of this basis is optimal to extract the cosmological information or not.  
Quantitative understanding of it is therefore the goal of this paper, and we will investigate it with the Fisher matrix analysis 
in Sec.~\ref{sec: results}.

\subsection{Alcock-Paczynski effect}
\label{sec:APeffect}
The observed IA power spectra of galaxies contain anisotropies induced not only from RSD and projection of the shape described in Sec.~\ref{sec:multipole power spectrum} but also from the Alcock-Paczynski (AP) effect \cite{1979Natur.281..358A}. The AP effect is caused by the apparent mismatch between underlying cosmology and fiducial model to convert the observed galaxy position, i.e., redshift and angular position on the sky, into the comoving space. 
As a result, the measured wave number perpendicular and parallel to the line of sight, $k_{\perp}$ and $k_{\parallel}$, are respectively modulated from true ones, $q_{\perp}$ and $q_{\parallel}$, as $k_{\perp}=(d_{\rm A}/d_{{\rm A}}^{{\rm fid}})q_{\perp}$ and $k_{\parallel}=(H/H^{{\rm fid}})^{-1} q_{\parallel}$, with $d_{\rm A}$ and $H$ being the angular diameter distance and Hubble parameter, respectively. The superscript ``fid'' implies quantities computed assuming the fiducial cosmology. As a result, the observed power spectrum with isotropic and anisotropic shifts due to the AP effect is given by \cite{2003ApJ...598..720S, 2020ApJ...891L..42T,PhysRevD.101.023510, 2022PhRvD.106d3523O}, 
\begin{equation}
P^{\rm XY, obs}(k_{\perp},k_{\parallel})=\frac{H(z)}{H^{{\rm fid}}(z)}\left\{\frac{d_{{\rm A}}^{{\rm fid}}(z)}{d_{{\rm A}}(z)}\right\}^2 P^{\rm XY}(q_{\perp},q_{\parallel}), \label{eq: def_PAP}
\end{equation}
where the superscript ``obs'' denotes observed quantities. 
Given the RSD and AP effect, we can simultaneously constrain the cosmologically informative parameters: the linear growth rate $f(z)$, the Hubble parameter $H(z)$ and the angular diameter distance $d_{\rm A}(z)$, using the power spectra~\cite{2011PhRvD..83j3527T,2014MNRAS.439.2515O,2012MNRAS.426.2719R}. 

\subsection{Gaussian covariance for the power spectrum multipoles}
\label{sec:covariance}
Assuming that our observables, galaxy density and ellipticity fields, follow the Gaussian distribution, we present the analytic covariance matrices of power spectrum multipoles expanded in terms of the normalized associated Legendre basis. This Gaussian assumption is reasonable as we consider the correlations at the large scales where linear theory is applied.
The Gaussian covariance matrices for the multipoles of the auto-power spectra of the galaxy density and ellipticity fields, $P^{\rm gg}$ and $P^{\rm EE}$, and their cross-power spectrum, $P^{\rm gE}$, are described as 
\begin{align}
&
{\rm COV}^{\rm XY, X'Y'}_{\ell m,\ell'm'}(k)=
\frac{2}{N_k}\int^{1}_{-1}{\rm d}\mu \Theta_{\ell, m}(\mu)
\Theta_{\ell',m'}(\mu)\notag \\
& \times \Bigl\{\widehat{P}^{\rm XX'}(k,\mu)\widehat{P}^{\rm YY'}(k,\mu)+\widehat{P}^{\rm XY'}(k,\mu)\widehat{P}^{\rm YX'}(k,\mu)\Bigr\} ~\label{eq:CovPlm},
\end{align}
where the superscripts ${\rm X}$, ${\rm Y}$, $\rm X'$, and $\rm Y'$ are the galaxy density field or the E-mode ellipticity field, respectively denoted as $\rm g$ and $\rm E$. 
The quantity $N_k=4 \pi k^2 \Delta {k} V_{\rm s}/(2 \pi)^3$ is the number of independent Fourier modes at the $k$ bin with width $\Delta k$ with $V_{\rm s}$ being the survey volume.  
We have added a hat on $P^{\rm XX'}$ to emphasize that  the auto-power spectra contain Poisson shot/shape noise contribution, namely $\widehat{P}^{\rm g g}=P^{\rm gg}+1/n_{\rm g}$ and $\widehat{P}^{\rm E E}=P^{\rm EE}+\sigma^2_{\rm \gamma}/n_{\rm g}$, where $n_{\rm g}$ and $\sigma_{\rm \gamma}$ are respectively the mean number density of galaxies and the scatter of the intrinsic shape. Since the cross power spectrum does not have it, we have, simply, $\widehat{P}^{\rm g E}=P^{\rm gE}$. 

The analytic expressions of the covariance matrix for the $\rm gE$ cross- and $\rm EE$ auto-power spectrum expanded in terms of the Legendre basis up to $\ell=4$ were already derived in Ref.~\cite{2021JCAP...11..061T}. 
As for the expansion in terms of the associated Legendre basis, Ref.~\cite{2023PhRvD.108h3533K} derived the covariance matrix for the $\rm gE$ power spectrum including the window and other observational effects. Here we newly derive the expression for the Gaussian covariance of the EE power spectrum expanded in terms of the normalized associated Legendre basis with $m>0$. Rewriting $\mu$ dependence of the full 2D power spectrum in Eq.~(\ref{eq:CovPlm}) to the normalized associated Legendre polynomials $\Theta_{\ell,m}(\mu)$, we perform the angular integral with the relation between three normalized associated Legendre polynomials and the Wigner 3-j symbols (e.g., \cite{MAVROMATIS1999101}),
\begin{align}
&
\Theta_{\ell_1,m_1}(\mu)\Theta_{\ell_2,m_2}(\mu)\Theta_{\ell_3,m_3}(\mu)
\notag \\&
=\sum_{L}\sum_{M}\sqrt{\frac{(2 \ell_1+1)(2\ell_2+1)(2 L+1)}{2}} 
\left(
\begin{array}{ccc}
\ell_1 & \ell_2& L\\
m_1 & m_2 & M
\end{array}
\right)
\notag \\ & \times
\left(
\begin{array}{ccc}
\ell_1 & \ell_2& L\\
0 & 0 & 0
\end{array}
\right) \Theta_{L,M} (\mu)\Theta_{\ell_{3},m_3}(\mu)~\label{eq:Gaunt},
\end{align}
where $L$ and $M$ are integers restricted by $L\geq|M|$ and the selection rules of the Wigner 3-j symbols.
Using the orthonormal property of $\Theta_{\ell,m}$ in Eq.~(\ref{eq:Gaunt}), we obtain the covariance for the $\rm gE$ cross- and $\rm EE$ auto-power spectrum multipoles. 
In Appendix~\ref{app: derivation cov}, starting from Eq.~(\ref{eq:CovPlm}), we provide the analytic expressions of the auto- and cross-covariance matrices for the gE and EE power spectrum multipoles. The first, second and third terms of Eqs.~(\ref{eq:COVgE})--(\ref{eq:COVEE}) correspond to the pure cosmic variance (CV $\times$ CV), the cross talk between the cosmic variance and Poisson noise (CV $\times$ P) and pure Poisson noise (P $\times$ P), respectively.
The explicit forms up to $\ell=4$ are given in Eqs.~(\ref{eq: COVP2222})--(\ref{eq: COVIIP4444}).

While Ref.~\cite{2021JCAP...11..061T} claimed that the multipole covariance matrices at $\ell > 4$ vanish,  each of the multipole covariances has all non-vanishing multipole contributions as Ref.~\cite{2016MNRAS.457.1577G} pointed out (see also Ref.~\cite{2020MNRAS.498L..77S}). 
As long as the selection rules in Wigner 3-j symbols in Eqs.~(\ref{eq:COVgE})--(\ref{eq:COVEE}) are satisfied, the CV $\times$ CV and CV $\times$ P terms can take non-zero values and thus have the cosmological information even at higher-order multipoles $\ell>4$ where the power spectrum multipoles themselves vanish. 
The P $\times$ P term also takes non-zero values at the higher-order multipoles although it is an isotropic quantity (see e.g., Refs.~\cite{2010PhRvD..82f3522T, 2021JCAP...11..061T,2023MNRAS.518.4976S}).
The higher-order multipole contributions ultimately have non-negligible impacts on the forecast results as we will discuss in Sec.~\ref{sec: results}. 

\section{Forecast formalism}
\label{sec: Forecast_formalism}

As mentioned in the previous section, even though a power spectrum multipole vanishes at a given order, its covariance does not necessarily vanish, indicating that the non-vanishing higher-order multipoles of the covariance contains the cosmological information.
In this section, we introduce the formalisms of the cumulative signal-to-noise ratio in Sec.~\ref{sec: SN} and the Fisher matrix in Sec.~\ref{sec: Fisher}, which will be used to quantify how much the cosmological information is contained in the higher-order multipoles of the covariance and how it depends on the choice of the basis for the multipole expansion in the next section.

\subsection{Signal-to-noise ratio}
\label{sec: SN}

Given the power spectrum multipoles $\widetilde{P}^{\rm XY}_{\ell,m}$ ($\rm X,Y= g \  or \ E$) and their covariance matrices, we can numerically calculate the cumulative signal-to-noise ratio via the following equation:
\begin{align}
\left(\frac{{S}}{{N}}\right)^2_{\ell_{\rm max}} 
&
=
\frac{V_{\rm s}}{4 \pi^2} \int^{k_{{\rm max}}}_{k_{{\rm min}}} k^2 d k \notag \\ &\times \sum^{\ell_{{\rm max}}}_{\ell,\ell'}\widetilde{P}^{\rm XY}_{\ell,m}(k) \{\widetilde{{\rm COV}}_{\ell m,\ell'm'}^{\rm XY,XY}(k)\}^{-1}\widetilde{P}^{\rm XY}_{\ell',m'}(k)
, \label{eq: def_SN_multipole}
\end{align}
where 
$k_{{\rm min}}$ and $k_{{\rm max}}$ are the minimum and maximum wave numbers used for the analysis, which will be specified later (see Sec.~\ref{sec: setup}), and $\widetilde{{\rm COV}}^{\rm XY,XY}_{\ell m,\ell' m'}(k)$ is related to Eq.~(\ref{eq:CovPlm}) through 
\begin{equation}
\widetilde{{\rm COV}}^{\rm XY,XY}_{\ell m,\ell' m'}(k)= \frac{N_{k}}{2} \times {\rm COV}^{\rm XY,XY}_{\ell m,\ell' m'}(k). \label{eq:reCovPlm}
\end{equation}

Even if the power spectrum multipole is zero at a given high $\ell$ ($\ell>4$ for our case at the linear theory limit), its covariance is not. Hence the vanishing multipoles would contribute to the non-vanishing ones at lower $\ell$ via the off-diagonal components of the covariance matrix.
As a result, the sum of the multipoles that contain linear-order contributions, up to $\ell=4$, does not necessarily coincide with the result obtained from the full 2D power spectrum. 
Regardless of whether the standard or associated Legendre basis is adopted, as $\ell_{{\rm max}}$ increases, the signal-to-noise ratio computed by Eq.~(\ref{eq: def_SN_multipole}) gradually approaches that obtained from the full 2D spectra given by
\begin{align}
&
\left(\frac{{S}}{{N}}\right)^2_{\rm 2D}
 =
\frac{V_{\rm s}}{4 \pi^2} \int^{k_{{\rm max}}}_{k_{{\rm min}}} k^2 d k \int^{1}_{-1} d \mu \notag \\
& \times
\Biggl[\frac{\{P^{\rm XY}(k,\mu)\}^2}{\widehat{P}^{\rm XX}(k,\mu)\widehat{P}^{\rm YY}(k,\mu)+\widehat{P}^{\rm XY}(k,\mu)\widehat{P}^{\rm YX}(k,\mu)}\Biggr]
. \label{eq: def_SN_2D}
\end{align}

When we consider only the shot/shape noise term (${\rm P} \times {\rm P}$) in the covariance matrix, which is the case where the cosmic variance is subdominant compared to the shot noise due to the limited number of samples, the signal-to-noise ratio for the multipole contributions [Eq.~(\ref{eq: def_SN_multipole})] has a simple form as
\begin{align}
\left(\frac{S}{N}\right)^2_{\ell_{\rm max}} 
=\frac{V_{\rm s}}{4 \pi^2} \int^{k_{{\rm max}}}_{k_{{\rm min}}} k^2 d k \frac{1}{N^{\rm X}N^{\rm Y}} \sum^{\ell_{\rm max}}_{\ell \geq m} \{\widetilde{P}^{\rm XY}_{\ell,m}(k)\}^2, \label{eq: SN_PP}
\end{align}
where $\rm X,Y= g \ or \ E$ and we used the orthonormal relation of $\Theta_{\ell,m}$. The shot/shape noise contribution, $N^{\rm X}$, indicates $N^{\rm g}=1/n_{\rm g}$ or $N^{\rm E}=\sigma^2_{\gamma}/n_{\rm g}$.

\subsection{Fisher matrix}
\label{sec: Fisher}

Ref.~\cite{2020ApJ...891L..42T} used the full 2D power spectra, $P(k,\mu)$, to demonstrate that combining the conventional galaxy clustering with the IA leads to improved constraints on the parameters $f(z),\ H(z)$ and $d_{\rm A}(z)$. 
Since the cosmological analysis of the observed $P(k,\mu)$ is complicated due to the large degree of freedom, it is more convenient to analyze the multipole moments. 
Thus, we extend the work of Ref.~\cite{2020ApJ...891L..42T} and perform the joint analysis of the multipoles of the galaxy density and ellipticity power spectra, taking into account the different expansion schemes. 
In this subsection, we provide the Fisher matrix formalism  \cite{1997PhRvL..79.3806T,1997ApJ...480...22T} for the joint analysis. 

Given the free parameters $\bm{\theta}=(b,f,b_{\rm K}, H/H^{{\rm fid}}, d_{\rm A}/d^{\rm  fid}_{\rm A})$ introduced in Sec.~\ref{sec:multipoles} and provided a set of observed power spectra ($P^{\rm gg}, \ P^{\rm gE}, \ P^{\rm EE}$), the Fisher matrix is evaluated with 
\begin{align}
&
F_{ij}^{\ell_{\rm max}}
=
\frac{V_{\rm s}}{4 \pi^2} \int^{k_{{\rm max}}}_{k_{{\rm min}}} k^2 d k \notag \\
& \times \sum_{a,b}^{N_{\rm P}}\sum^{\ell_{{\rm max}}}_{\ell,\ell'}\frac{\partial \widetilde{P}^{a}_{\ell
,(m,m',m'')
}(k)}{\partial \theta_i}\{\widetilde{{\rm COV}}^{-1}\}^{a,b}_{\ell,\ell'}\frac{\partial \widetilde{P}^{b}_{\ell',(m,m',m'')}(k)}{\partial \theta_j}
, \label{eq: def_Fisher_multipole}
\end{align}
where $N_{\rm P}$ is the number of different power spectra and 
$\widetilde{P}^{a}_{\ell,{(m,m',m'')}}(k)=$   $\{\widetilde{P}^{\rm gg}_{\ell,m}(k), \ \widetilde{P}^{\rm gE}_{\ell,m'}(k), \ \widetilde{P}^{\rm EE}_{\ell,m''}(k)\}$ for $a=1,2,3$, respectively, 
where the lower limit for the summation of $\ell$ for each power spectrum should be a chosen value of $m, m'$, and $m''$. 
As shown in Appendix~\ref{app: dev_power_spectra}, derivatives of the power spectra become non zero up to $\ell=6$ at the linear theory limit due to the AP effect.

When combining all the statistics, the covariance matrix is given by 
\begin{equation}
\widetilde{{\rm COV}}_{\ell,\ell'}^{a,b}=
\begin{pmatrix}
{\widetilde{\rm{COV}}^{\rm gg,gg}_{\ell m,\ell' m}} & {\widetilde{\rm{COV}}^{\rm gg,gE}_{\ell m,\ell' m'}} & {\widetilde{\rm{COV}}^{\rm gg,EE}_{\ell m,\ell' m''}}\\
{\widetilde{\rm{COV}}^{\rm gE,gg}_{\ell m',\ell' m}} & {\widetilde{\rm{COV}}^{\rm gE,gE}_{\ell m',\ell' m'}} &
{\widetilde{\rm{COV}}^{\rm gE,EE}_{\ell m',\ell' m''}} \\
{\widetilde{\rm{COV}}^{\rm EE,gg}_{\ell m'',\ell' m}} &
{\widetilde{\rm{COV}}^{\rm EE,gE}_{\ell m'',\ell' m'}} &
{\widetilde{\rm{COV}}^{\rm EE,EE}_{\ell m'',\ell' m''}}
\end{pmatrix}, \label{eq: def_all_cov}
\end{equation}
where the elements in the parentheses are calculated through Eq.~(\ref{eq:reCovPlm}). On the other hand, the Fisher matrix for the full 2D case is given by Refs.~\cite{2020ApJ...891L..42T,2022PhRvD.106d3523O}, 
\begin{align}
F_{ij}^{\rm 2D}
& =
\frac{V_{\rm s}}{4 \pi^2} \int^{k_{{\rm max}}}_{k_{{\rm min}}} k^2 d k \int^{1}_{-1} d \mu \notag \\
& \times
 \sum_{a,b}^{N_{\rm P}} \frac{\partial P^{a}(k,\mu)}{\partial \theta_i} \{\widetilde{{\rm COV}}^{-1}\}^{a,b}\frac{\partial P^{b}(k,\mu)}{\partial \theta_j}
. \label{eq: def_Fisher_2D}
\end{align}
Here, we construct the covariance matrix $\widetilde{{\rm COV}}^{a,b}$ in the same way as Eq.~(\ref{eq: def_all_cov}) but with the elements using the full 2D spectra, 
${\widetilde{\rm{COV}}^{\rm XY,X'Y'}}=\widehat{P}^{\rm XX'}\widehat{P}^{\rm YY'}+\widehat{P}^{\rm XY'}\widehat{P}^{\rm YX'}$. As in the signal-to-noise ratio, the parameter constraints obtained from the multipoles get tighter and approach the full 2D result as $\ell_{\rm max}$ increases.

Given the Fisher matrix, we can obtain the two-dimensional error ellipses for a specific pair of parameters from $\bm{\theta}=(b,f,b_{\rm K}, H/H^{{\rm fid}}, d_{\rm A}/d^{\rm fid}_{\rm A})$ by constructing the $2\times 2$ submatrix from the inverse Fisher matrix, $\left(F^{-1} \right)_{ij}$, while the other parameters are marginalized over. 
Also the one-dimensional marginalized error on the $i$th parameter $\theta_i$ is obtained by the $i$th diagonal component, $\sigma_{\theta}=\sqrt{\left(F^{-1} \right)_{ij}}$.
\begin{figure*}
\centering
\includegraphics[width=0.9\textwidth]{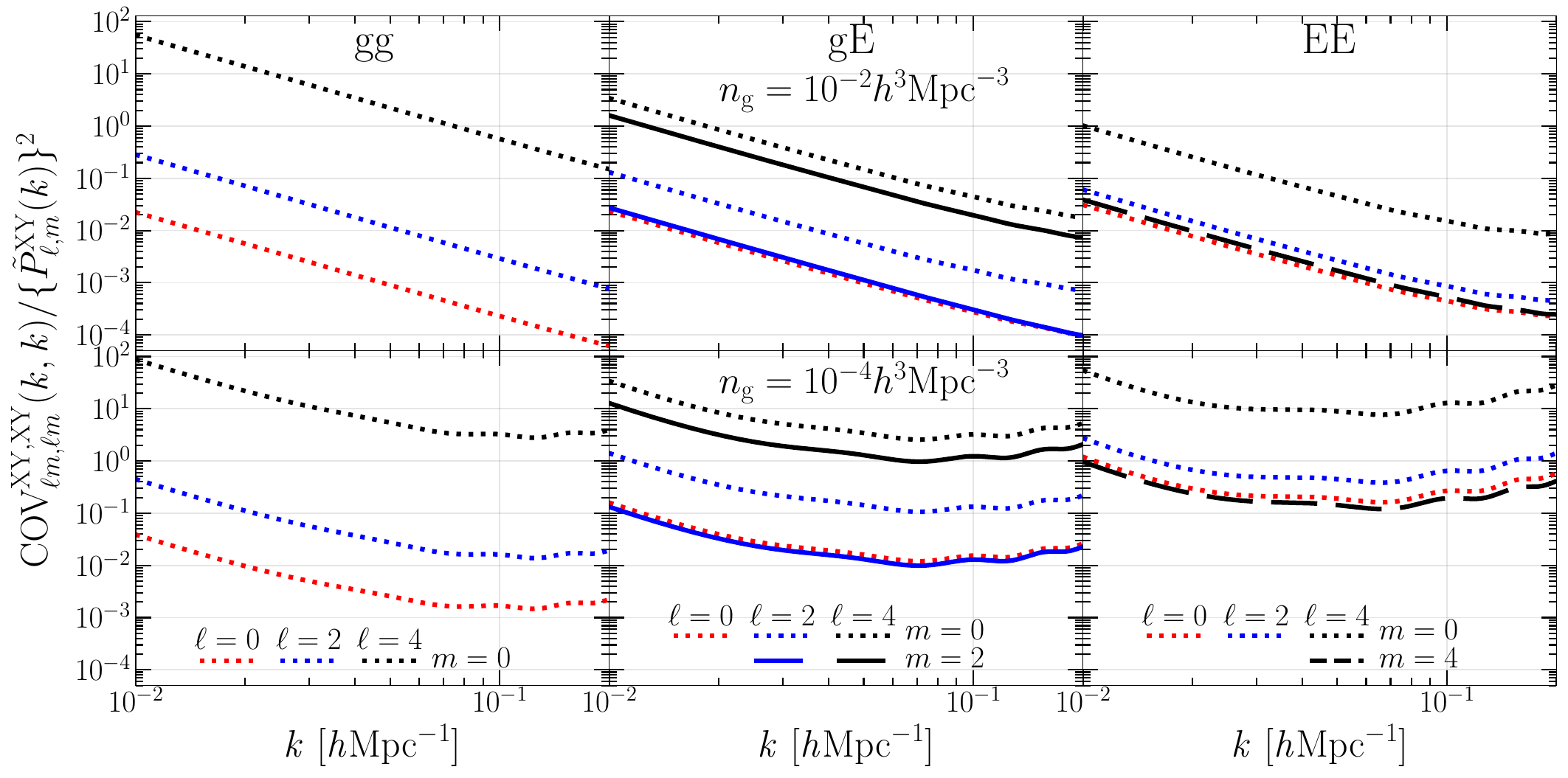}
\caption{Diagonal components of the covariance matrix normalized by the survey volume, divided by the square of each power spectrum multipole, ${\rm COV}^{\rm XY, XY}_{\ell m,\ell m}(k,k)(V_{\rm s}/[{ h^{-3}{\rm Gpc}^3}])/\{\widetilde{P}^{\rm XY}_{\ell,m}(k)\}^2$, where ${\rm XY}=\{ {\rm gg}, {\rm gE}, {\rm EE} \}$ from the left to right panels, respectively.
The results for the high- and low-number density samples are shown in the upper and lower rows, respectively. 
The $\ell=0$, $2$, and $4$ moments are colored by red, blue, and black, and the moments with $m=0$, $2$, and $4$ are shown by the dotted, solid, and dashed lines, respectively. }
\label{fig:  Cov_P_3}
\end{figure*}
\section{Results}
\label{sec: results}
In this section, we quantitatively show the information content encoded in the multipoles of the IA power spectra in the application of the different multipole expansion scheme.
After briefly providing our setup for the numerical calculation in Sec.~\ref{sec: setup}, we first show the fractional errors of the power spectrum multipoles in Sec.~\ref{sec: fractional errors} using the formulas presented in Sec.~\ref{sec:multipoles}. We then show the signal-to-noise ratios in Sec.~\ref{sec: detectability} and the geometric and dynamical constraints expected from the multipoles of the individual power spectra in Sec.~\ref{sec: parameter constraints}.
We also present well-known results of the galaxy auto-power spectrum, $P^{\rm gg}$, as a reference. 

\subsection{Setup}
\label{sec: setup}
We consider two hypothetical galaxy surveys with the different number densities, $n_{\rm g}=10^{-2} \ h^3{\rm Mpc}^{-3}$ and $n_{\rm g}=10^{-4} \ h^3{\rm Mpc}^{-3}$ but the other parameters being the same. The former one roughly corresponds to the Bright Galaxy Survey (BGS) samples observed by Dark Energy Spectroscopic Instrument (DESI) \cite{2016arXiv161100036D}, while the latter to galaxy samples such as the Luminous Red Galaxies (LRG) and the Emission Line Galaxies (ELG). For each sample, we fix the redshift and the galaxy bias to $z=0.5$, and $b=1$, respectively. Following the work of Ref.~\cite{2022PhRvD.106d3523O}, we set the amplitude of IA, $A_{\rm IA}$, to $A_{\rm IA}=18$, assuming that we can estimate the IA of the dark matter halos that host the above galaxies \cite{Shi_2021}.
We adopt a value of the scatter of the intrinsic shape considered in up-coming galaxy surveys, $\sigma_{\gamma}=0.2$.  
Since our formalism is based on the linear theory, to be conservative, we restrict our analysis to large scales and set $k_{{\rm min}}=2\pi/V_{\rm s}^{1/3}$ and $k_{{\rm max}}= 0.1 \ h {\rm Mpc}^{-1}$. Our conservative choice of $k_{\rm max}$ can justify the use of linear expressions and Gaussian covariance \cite{Takahashi_2009, Takahashi_2010,Blot_2014,Blot_2019,Chan_2017}.
All numerical results in following subsections are normalized by the survey volume. 
Except for the results in Sec.~\ref{sec: fractional errors}, in this section, we focus on the results expected from the high number density sample and discuss the results for the low number density sample in Sec.~\ref{sec: discussion}.
\begin{figure*}
\includegraphics[width=0.49\textwidth]{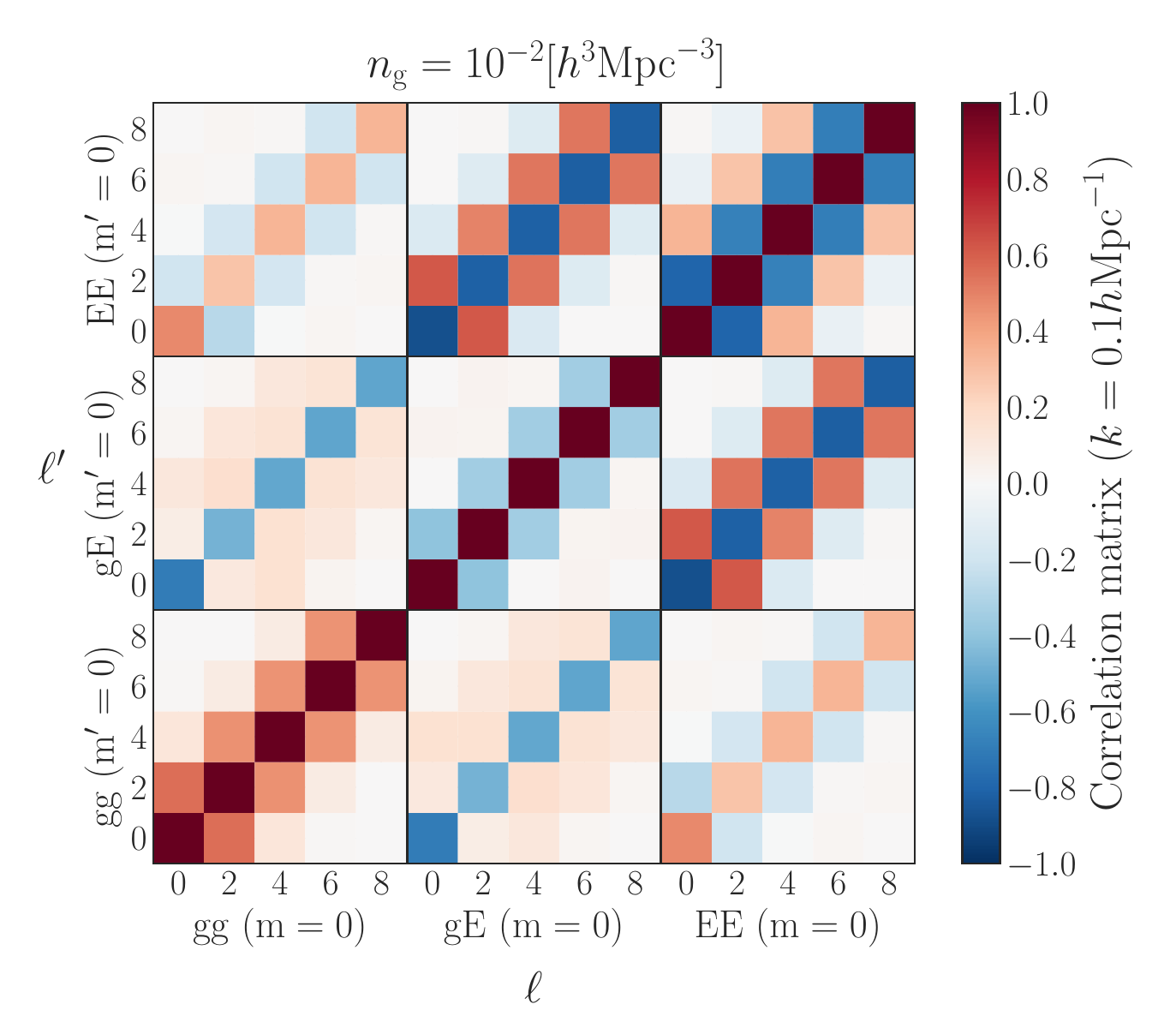}
\hfill
\includegraphics[width=0.49\textwidth]{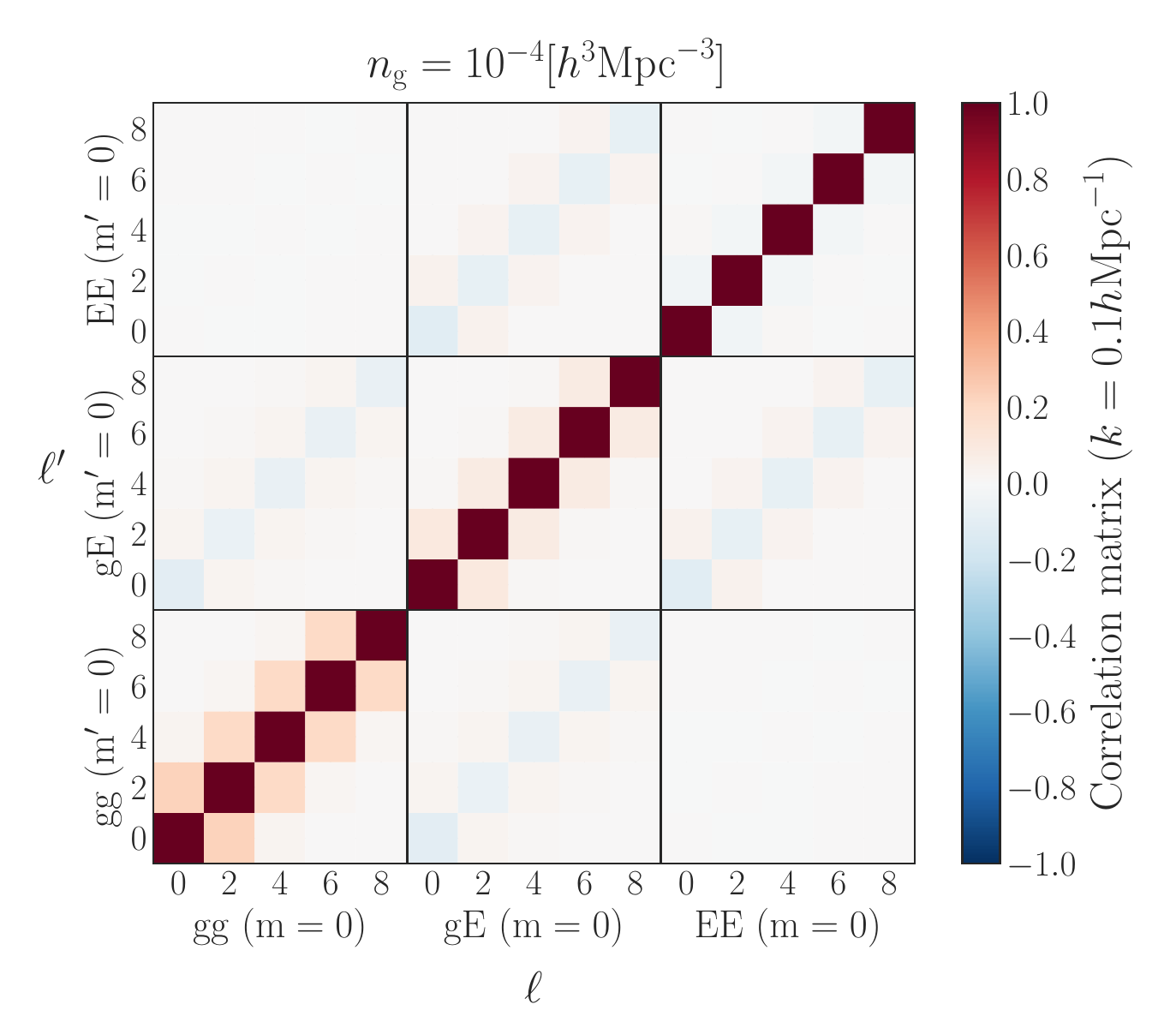}
\vfill
\includegraphics[width=0.49\textwidth]{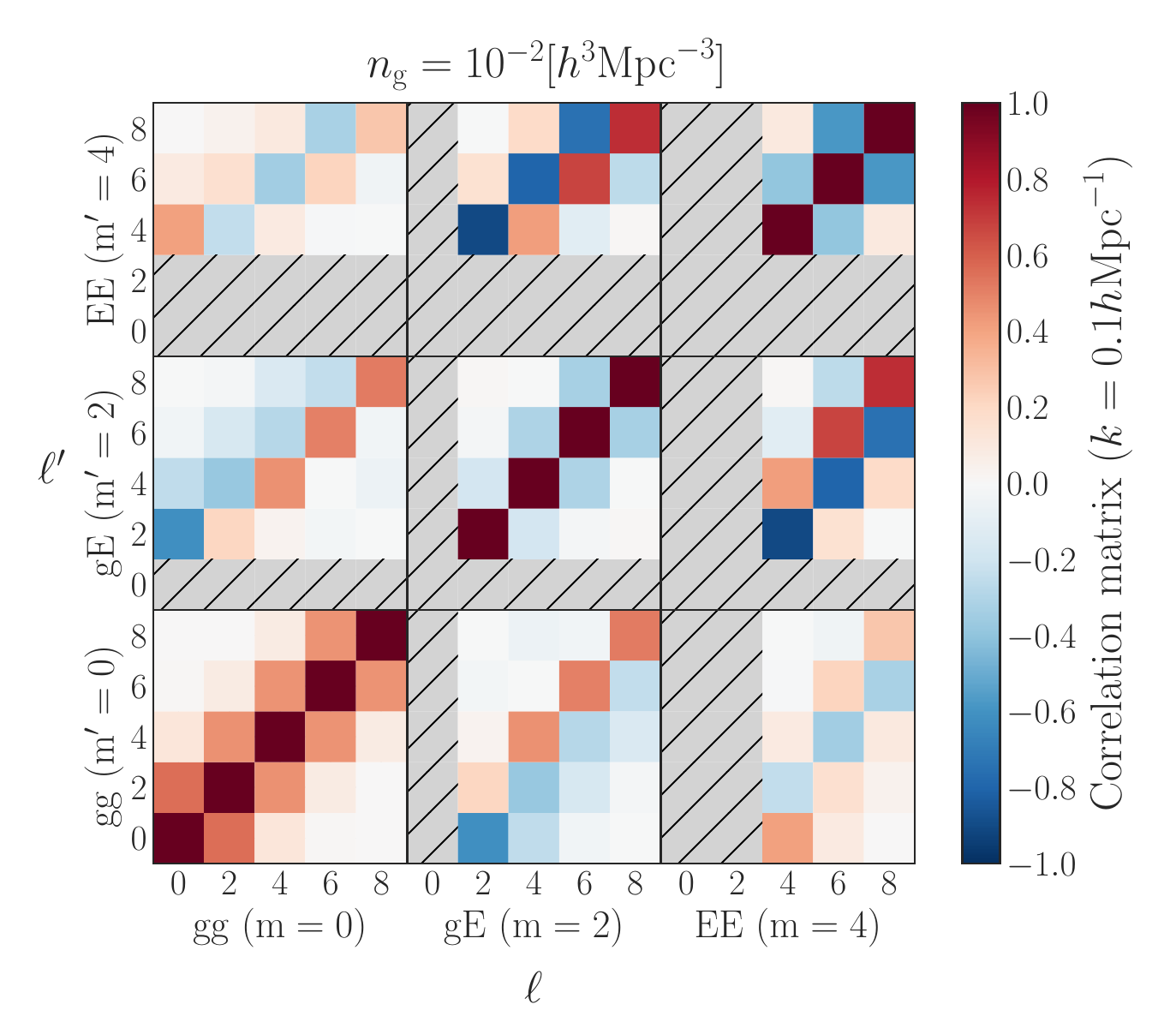}
\hfill
\includegraphics[width=0.49\textwidth]{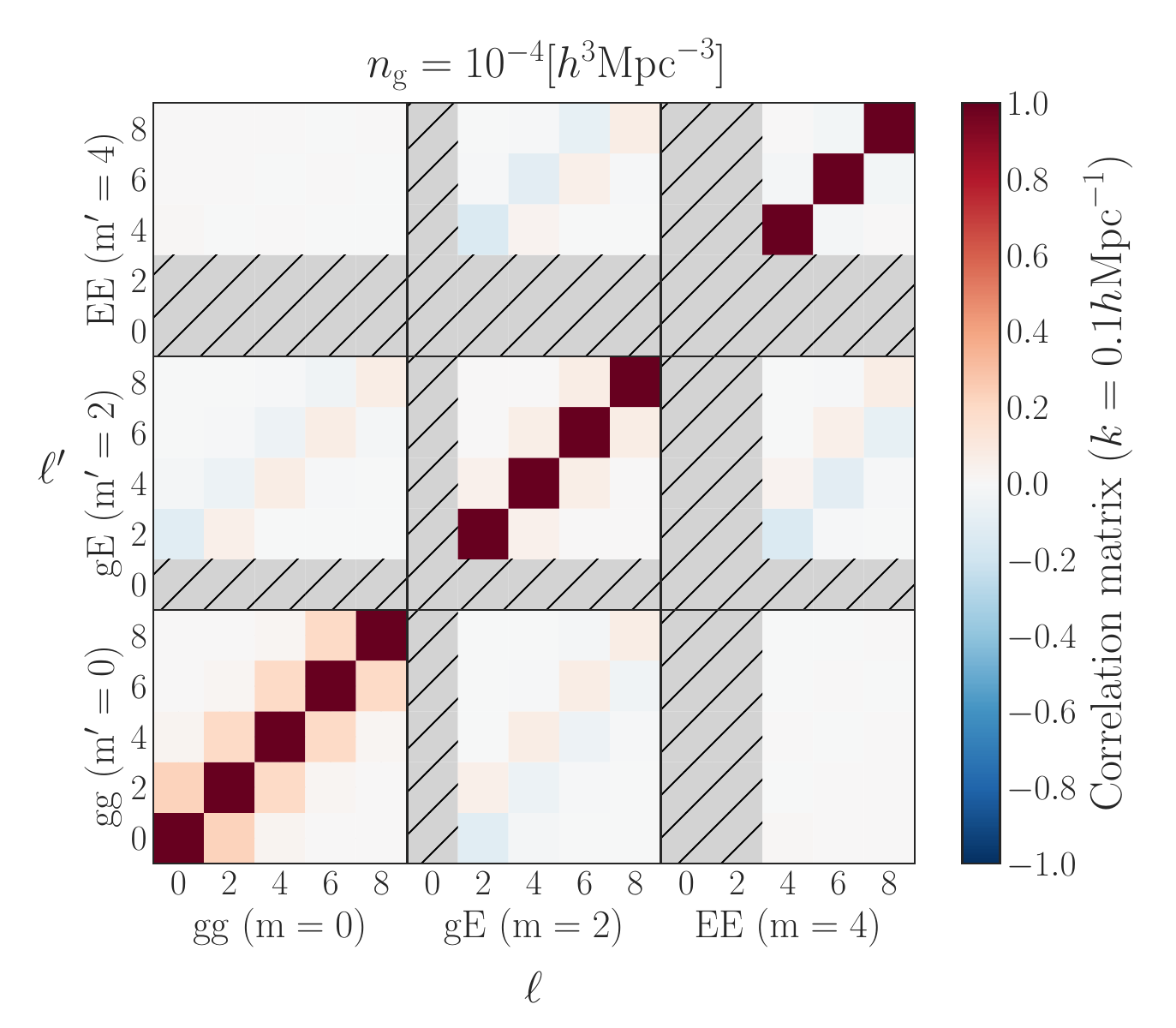}
\caption{Correlation matrix obtained from the auto- and cross-power spectrum multipoles at $k=0.1 h{\rm Mpc}^{-1}$. Here, we show the multipole coefficients, $\ell$ and $\ell'$, up to $8$. The upper and lower panels show the results obtained from the multipoles of the IA spectra for $m=0$ and $m>0$, respectively. The left and right panels show the results obtained from the high and low number density cases, respectively. Note that the hatched areas filled in gray indicate regions where no values are present ($\ell \geq m$ and $\ell' \geq m'$).}
\label{fig:COVl8}
\end{figure*}
\subsection{Error of power spectrum multipoles}
\label{sec: fractional errors}

To compare the information content of a multipole moment of the IA power spectra in the associated Legendre basis with those in the conventional Legendre one, we first look at the fractional errors, which are defined by the diagonal components of the covariance matrix divided by the square of the power spectrum multipoles. 

The left to right panels of Fig.\ref{fig: Cov_P_3} show the fractional errors of the multipoles of $P^{\rm gg}$, $P^{\rm gE}$ and $P^{\rm EE}$, respectively. 
The top and bottom panels show the results for the high and low number density samples, respectively. In the low number density case, turnover scales arise due to the shot noise contribution. This effect is particularly evident in the EE power spectrum, since the P $\times$ P contribution dominates the noise. Consequently, the shapes of the fractional errors are influenced by the shapes of the power spectrum multipoles.
We find that the fractional errors on $\ell=2$ and $4$ moments of the associated Legendre basis ($m>0$) are lower than those of the standard Legendre one ($m=0$) for the IA power spectra, $P^{\rm gE}$ and $P^{\rm EE}$ in both of the two different number densities. This is because the information is compressed into higher-order multipole moments, leading to the minimum set of multipoles in the associated Legendre basis.
Comparing the lowest-order multipoles of the IA power spectra between the two different bases, the results with $m>0$ have slightly lower fractional errors in the low number density samples and vice versa in the high number density ones. 
That is, a beneficial basis to extract more information from a single multipole moment depends on the sample under consideration. 
Although the multipole of $P^{\rm EE}$ in the associated Legendre basis is expressed as the single multipole, namely $\widetilde{P}^{\rm EE}_{4,4}$, it does not necessarily give the lowest fractional errors as shown in the upper-right panel. This would reflect the fact that the information from the higher-order multipoles in the covariance matrix is not negligible, as we discuss in detail in Sec.~\ref{sec: detectability}.

To see the correlations between the power spectrum multipoles, we show the correlation matrix, ${\rm COV}^{\rm XY,X'Y'}_{\ell m,\ell' m'}/\sqrt{{\rm COV}^{\rm XY,XY}_{\ell m,\ell m} \times {\rm COV}^{\rm X'Y',X'Y'}_{\ell' m',\ell' m'}}$, as a function of $\ell$ and $\ell'$ fixing the wavenumber to $k=0.1h{\rm Mpc}^{-1}$ in Fig.~\ref{fig:COVl8}. The left and right panels show the results for the high and low number density samples, respectively. We show the results obtained from the multipoles of the IA power spectra for $m=0$ and $m>0$, which are presented in the upper and lower panels, respectively. Note that the correlations still appear even at $\ell>4$ where the power spectrum multipoles vanish, as discussed above. For the high number density case, the contribution of the shot noise has less impact, i.e., cosmic variance dominates, leading to the stronger correlations between statistics that trace the same underlying matter density field. Since the sign of the multipoles in the IA power spectra at a given order varies with the basis choice, the sign of the covariance matrix varies accordingly. Interestingly, we find that the cross-correlation between IA power spectra, $P^{\rm gE}$ and $P^{\rm EE}$, shows a stronger correlation compared to the correlation between $P^{\rm gg}$ and IA power spectra, as clearly shown in the high number density case.

\begin{figure*}
\centering
\includegraphics[width=0.9\textwidth]{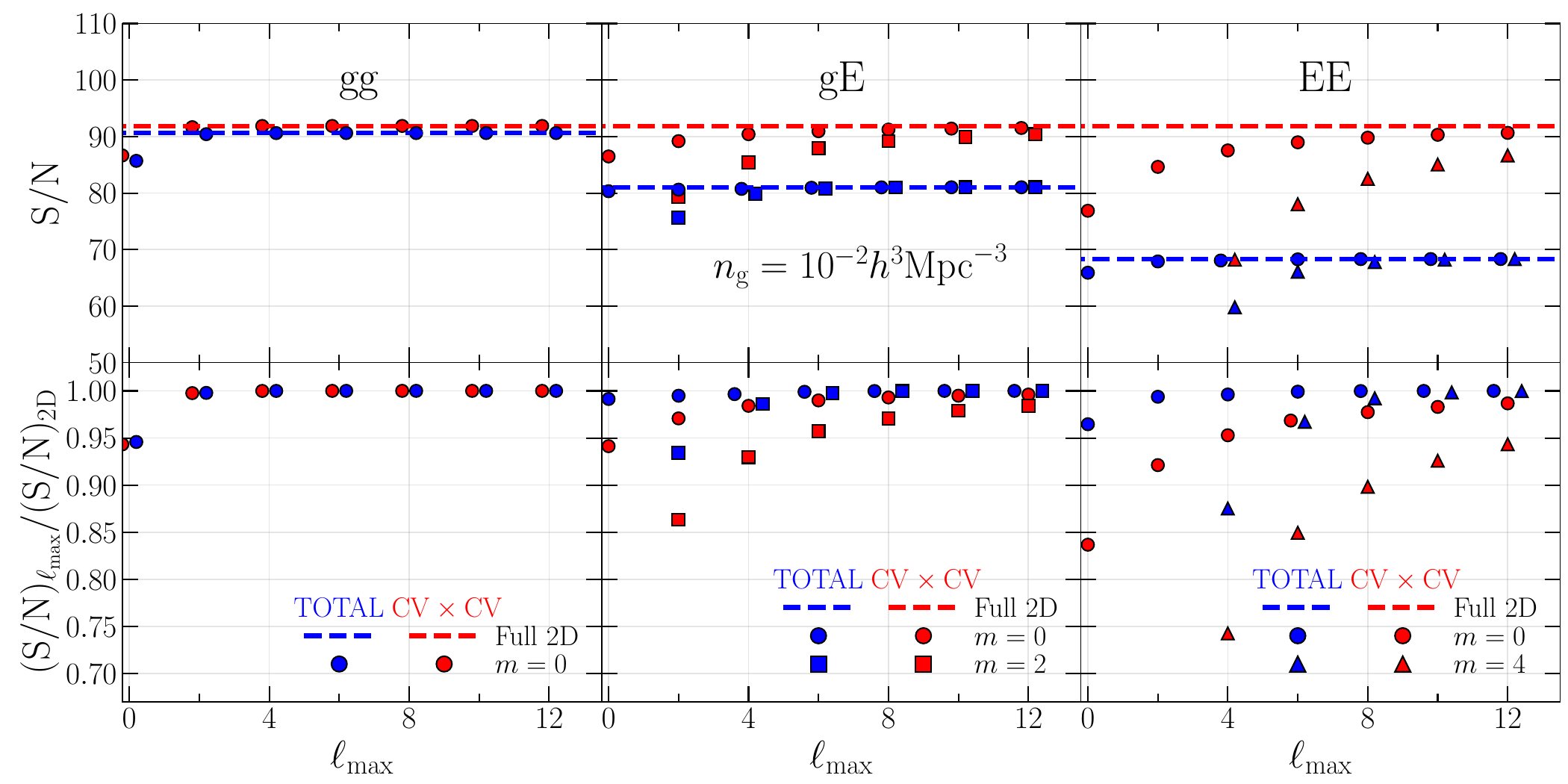}
\caption{{\it Top panels}: signal-to-noise ratios normalized by the survey volume, $S/N(V_{\rm s}/[h^{-3}{\rm Gpc^3}])^{-1/2}$, for different choices of $m$ in $\widetilde{P}^{\rm XY}_{\ell,m}(k)$ expected from the high number density sample. From the left to right panels, the results are obtained from $P^{\rm gg}$, $P^{\rm gE}$, and $P^{\rm EE}$, respectively. 
The circles show the multipoles in the standard Legendre basis as a function of $\ell_{\rm max}$.
The squares and triangles show those in the associated Legendre basis.
They are compared with the results from the full 2D spectra shown by the dashed horizontal lines. 
The red and blue colors indicate the results given with the noise from the cosmic variance only (CV $\times$ CV) and total noise contributions (CV $\times$ CV + CV $\times$ P + P $\times$ P), respectively. {\it Bottom panels}: ratio of the signal-to-noise ratios obtained from the multipole and full 2D spectra, namely $(S/N)_{\ell_{\rm max}}/(S/N)_{\rm 2D}$.
Note that some symbols are slightly shifted horizontally from integer values for visibility.
}
\label{fig:SN}
\end{figure*} 
\subsection{Signal-to-noise ratio}
\label{sec: detectability}

Here we estimate the signal-to-noise ratio obtained from the multipoles and full 2D spectra and see how the contributions of the multipoles in the two different bases reach the full 2D information by increasing $\ell_{\rm max}$.
The top panels in Fig.~\ref{fig:SN} show the signal-to-noise ratios of each of power spectra, $P^{\rm gg}$, $P^{\rm gE}$, and $P^{\rm EE}$.
We show the results obtained from the multipole spectra as a function of $\ell_{\rm max}$ as well as those from the full 2D spectra. 
The results including all the noise contributions, namely both the cosmic variance and the Poisson noise and their crosstalk (CV $\times$ CV $+$ CV $\times$ P $+$ P$\times$ P), are colored by blue. Those with the noise from the cosmic variance only are colored by red.
The bottom panels show the ratio of the signal-to-noise ratios between the multipole and 2D spectra, $(S/N)_{\ell_{\rm max}}/(S/N)_{\rm 2D}$.
The information encoded in the multipoles of $P^{\rm gg}$ becomes almost equivalent to that in the full 2D case when summing up contributions up to $\ell=4$ in the Legendre basis, consistent with the previous work \cite{2011PhRvD..83j3527T}.
When we consider only the ${\rm CV} \times {\rm CV}$ term in the covariance, the signal-to-noise ratios of multipoles for the IA spectra in both of the two different bases slowly converge to those in the full 2D case compared with $P^{\rm gg}$, indicating the non-negligible contributions from higher-order multipoles. This behavior is more significant in the auto-power spectrum, $P^{\rm EE}$, than the cross-power spectrum $P^{\rm gE}$.
We also find an interesting feature that the results for the associated Legendre basis converge more slowly than those for the conventional Legendre one. 
Note that all the full 2D spectra (i.e., $P^{\rm gg}$, $P^{\rm gE}$ and $P^{\rm EE}$) have the same signal-to-noise ratios in the case of ${\rm CV} \times {\rm CV}$ since their information is originated from the same underlying matter density field.

As described in Sec.~\ref{sec: SN}, when we consider only the shot/shape noise (${\rm P} \times {\rm P}$), 
the signal-to-noise ratio for the power spectra can have a simple form [see Eq.~(\ref{eq: SN_PP})].
In this specific case, the ratio of the signal-to-noise ratios obtained from the multipoles at a given $\ell_{\rm max}$ and the full 2D spectra, $\left[\left(S/N\right)^2_{\ell_{\rm max}}/\left(S/N\right)^2_{\rm 2D}\right]$, is further simplified using the relation derived in Appendix \ref{app: mathematical relations} as follows:
\begin{align}
\frac{\left(S/N\right)^2_{\ell_{\rm max}}}{\left(S/N\right)^2_{\rm 2D}}
=\frac{\sum^{\ell_{\rm max}}_{\ell \geq m} \left(Q^{\rm XY}_{\ell,m}\right)^2}{\int^{1}_{-1}d \mu \{Q^{\rm XY}(\mu)\}^2}
=\frac{\sum^{\ell_{\rm max}}_{\ell \geq m} \left(Q^{\rm XY}_{\ell,m}\right)^2}{\sum^{\bar\ell_{\rm max}}_{\ell \geq m} \left(Q^{\rm XY}_{\ell,m}\right)^2}, \label{eq: SN_PP ratio}
\end{align}
where $Q^{\rm XY}(\mu)\equiv P^{\rm XY}(k,\mu)/P_{\rm L}(k)$ and $Q^{\rm XY}_{\ell,m}\equiv\widetilde{P}^{\rm XY}_{\ell,m}(k)/P_{\rm L}(k)$.
The integer $\bar\ell_{\rm max}$ is the highest power of $\mu$ in the function $P^{\rm XY}(k,\mu)$. Namely, $\bar\ell_{\rm max}=4$ for our case at the linear theory limit. 
Eq.~(\ref{eq: SN_PP ratio}) leads to $\left(S/N\right)_{\ell_{\rm max}}=\left(S/N\right)_{\rm 2D}$ when $\ell_{\rm max}\geq \bar\ell_{\rm max}$ for arbitrary $m$. 
Thus, irrespective of the basis we choose, we can extract the full 2D information using the multipoles only up to $\ell=4$ for the ${\rm P} \times {\rm P}$ only case, unlike the ${\rm CV} \times {\rm CV}$ only case.

\begin{figure*}
\centering
\includegraphics[width=0.9\textwidth]{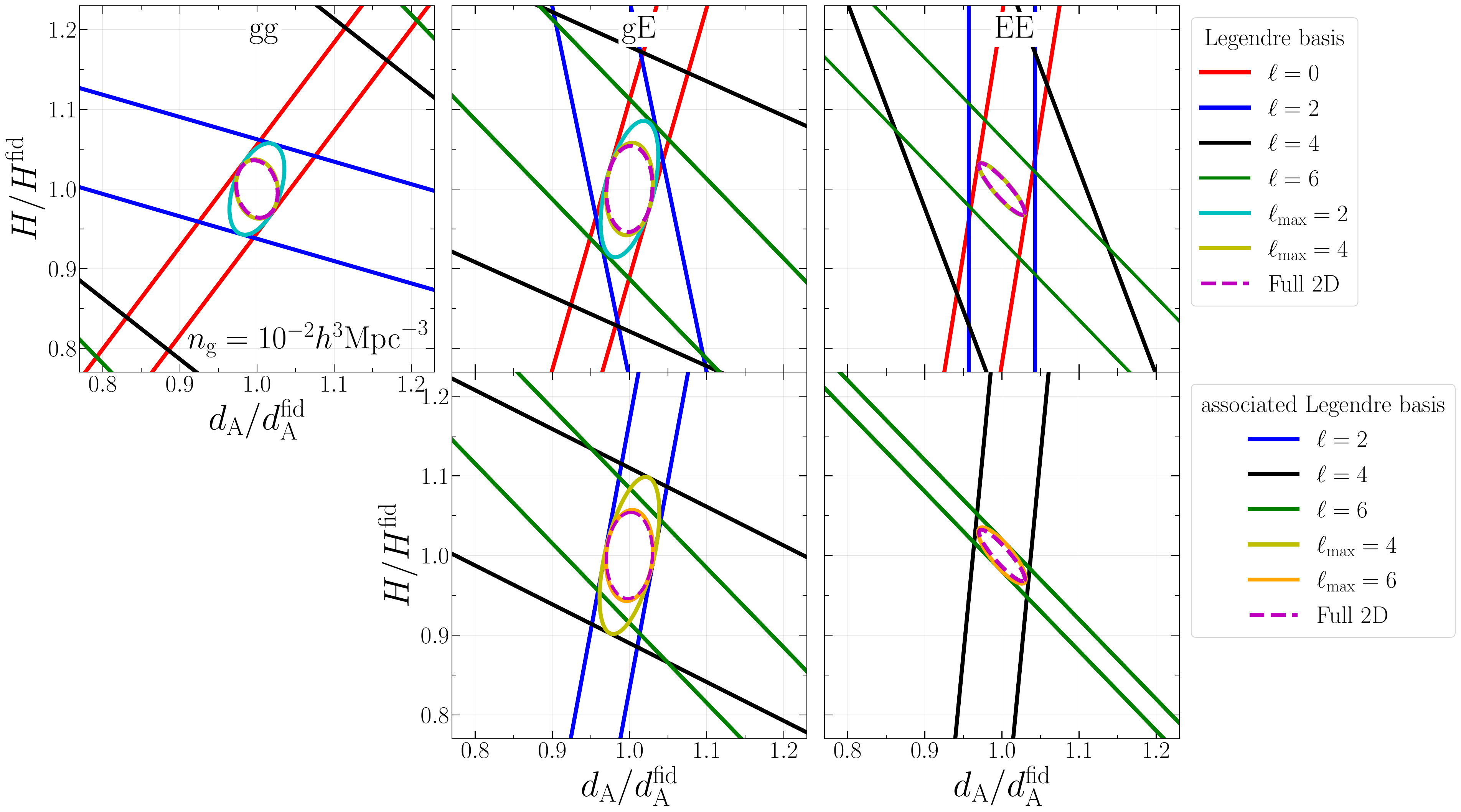}
\caption{
{\it Top panels}: two-dimensional error contours (68 \% C.L.) on the geometric distances, $d_{\rm A}$ and $H$, expected from the high number density sample. The solid contours from the left to right panels respectively show the results obtained from multipole contributions of $P^{\rm gg}_{\ell,m}$, $P^{\rm gE}_{\ell,m}$, and $P^{\rm EE}_{\ell,m}$ in the Legendre basis ($m=0$). The red, blue, black, and green contours show the results from individual multipoles, while the cyan, yellow, and orange ones shows the results of the combinations up to $\ell_{\rm max}$. 
For reference, results from the full 2D spectra are shown by the dashed contours. {\it bottom panels}: Same as the top panels but the results of $P^{\rm gE}$ and $P^{\rm EE}$ in the associated Legendre basis, $m=2$ and $m=4$, respectively. }
\label{fig:  2D_error}
\end{figure*}
\begin{figure*}
\centering
\includegraphics[width=0.9\textwidth]{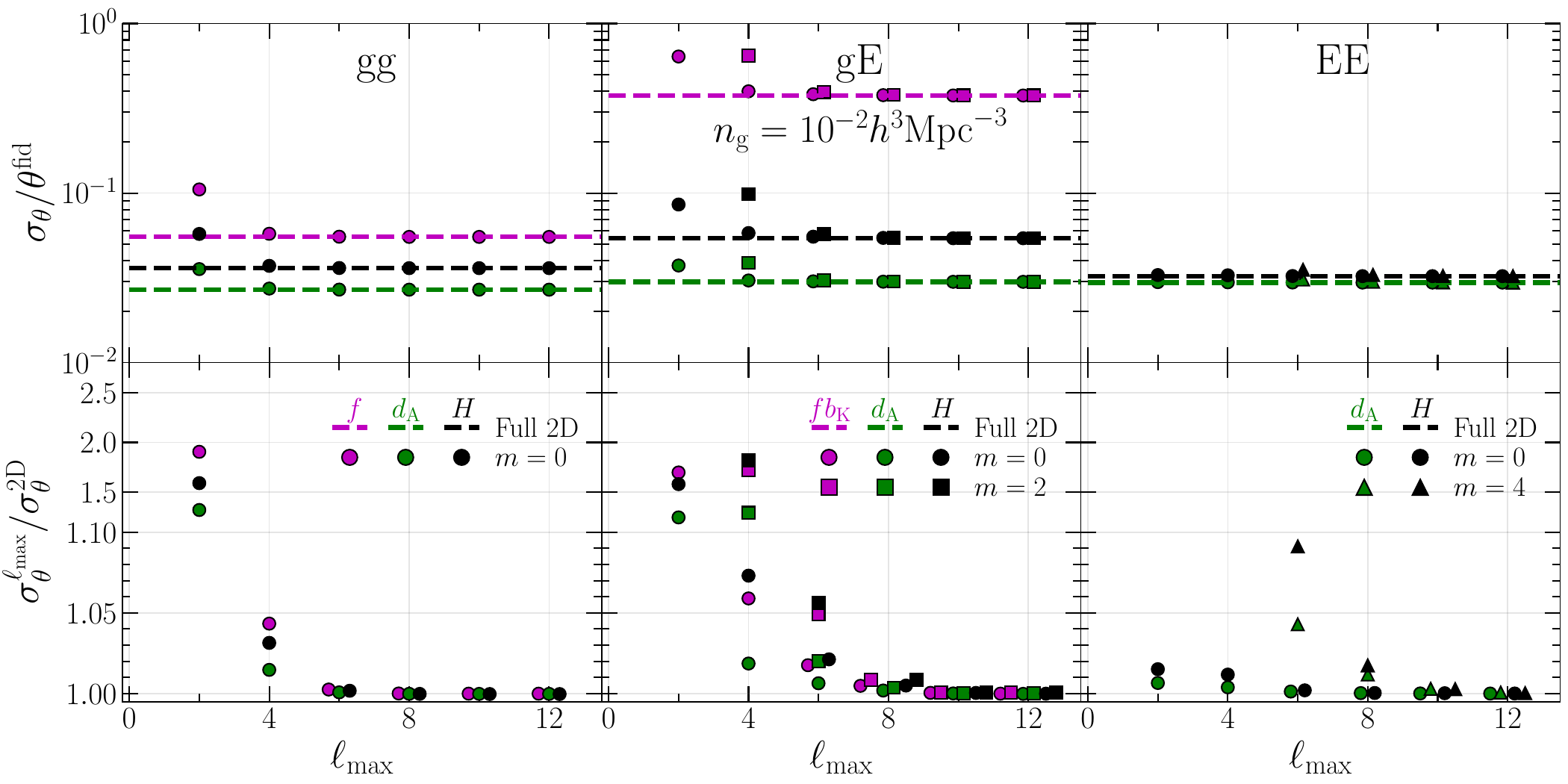}
\caption{{\it Top panels}: one-dimensional fractional errors on the growth rate, $f$ or $fb_{\rm K}$ (magenta), and the geometric distances, $d_{\rm A}$ (green) and $H$ (black), normalized by the square root of the survey volume, $\sigma_{\theta}(V_{\rm s}/[h^{-3}{\rm Gpc^3}])^{1/2}/\theta^{\rm fid}$, expected from the high number density sample. From the left to right panels, the results are obtained from $P^{\rm gg}$, $P^{\rm gE}$, and $P^{\rm EE}$, respectively. The symbols and dashed horizontal lines are the same as in Fig.\ref{fig:SN}. {\it Bottom panels}: ratio of the one-dimensional errors from the multipoles and full 2D spectra, $\sigma_{\theta}^{\ell_{\rm max}}/\sigma_{\theta}^{\rm 2D}$. Note that the vertical axis mixes different linear scalings. 
}
\label{fig:Fisher1D_figures}
\end{figure*}

Hence, considering the total noise contributions in the signal-to-noise ratio for the high number density sample where the cosmic variance dominates, higher-order multipoles such as $\ell=6$ or $8$ must be used to obtain the information comparable to the full 2D case particularly when the associated Legendre basis is adopted.

\begin{figure*}[t]
\centering
\includegraphics[width=0.9\textwidth]{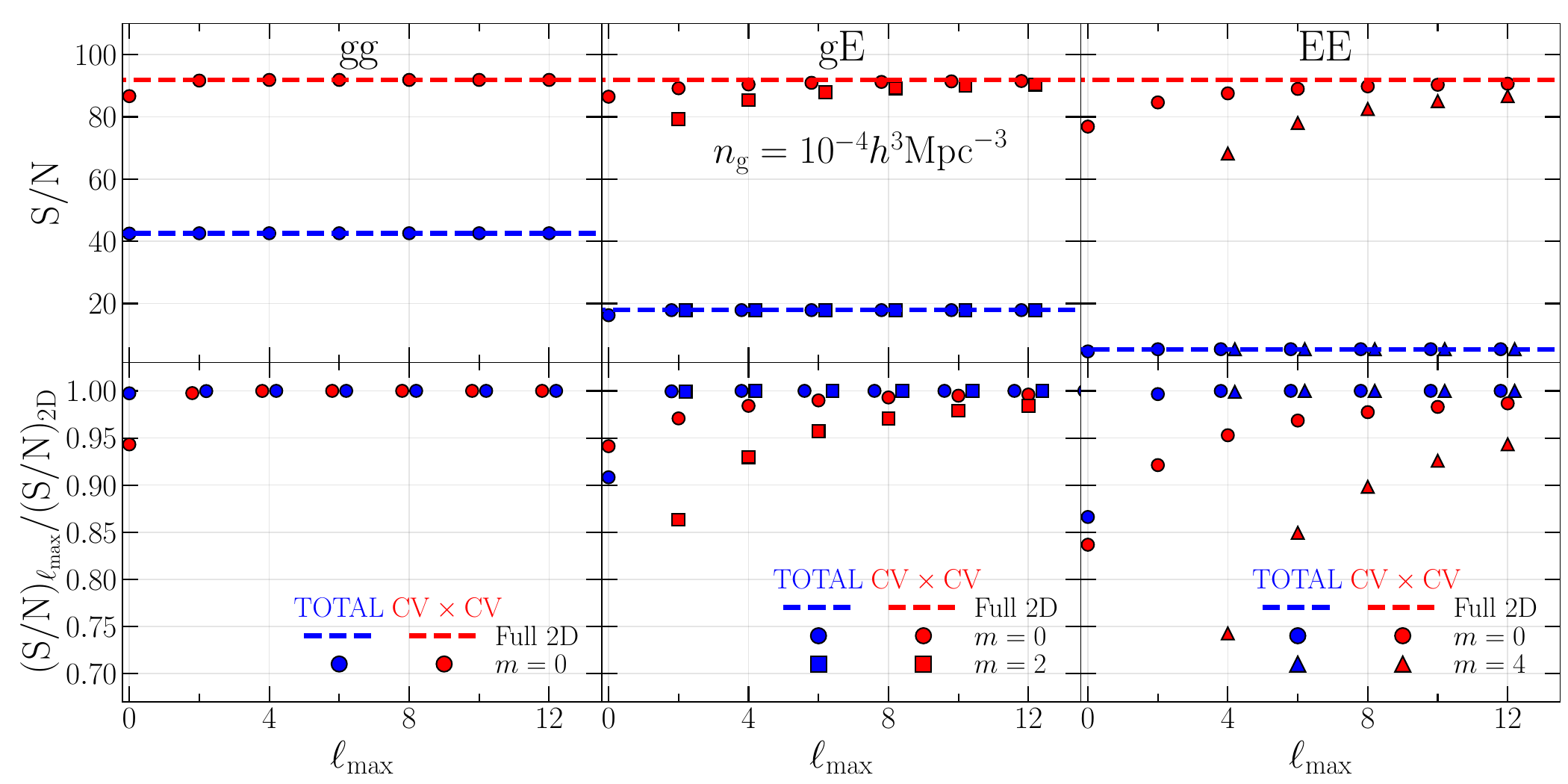}
\vfill
\includegraphics[width=0.9\textwidth]{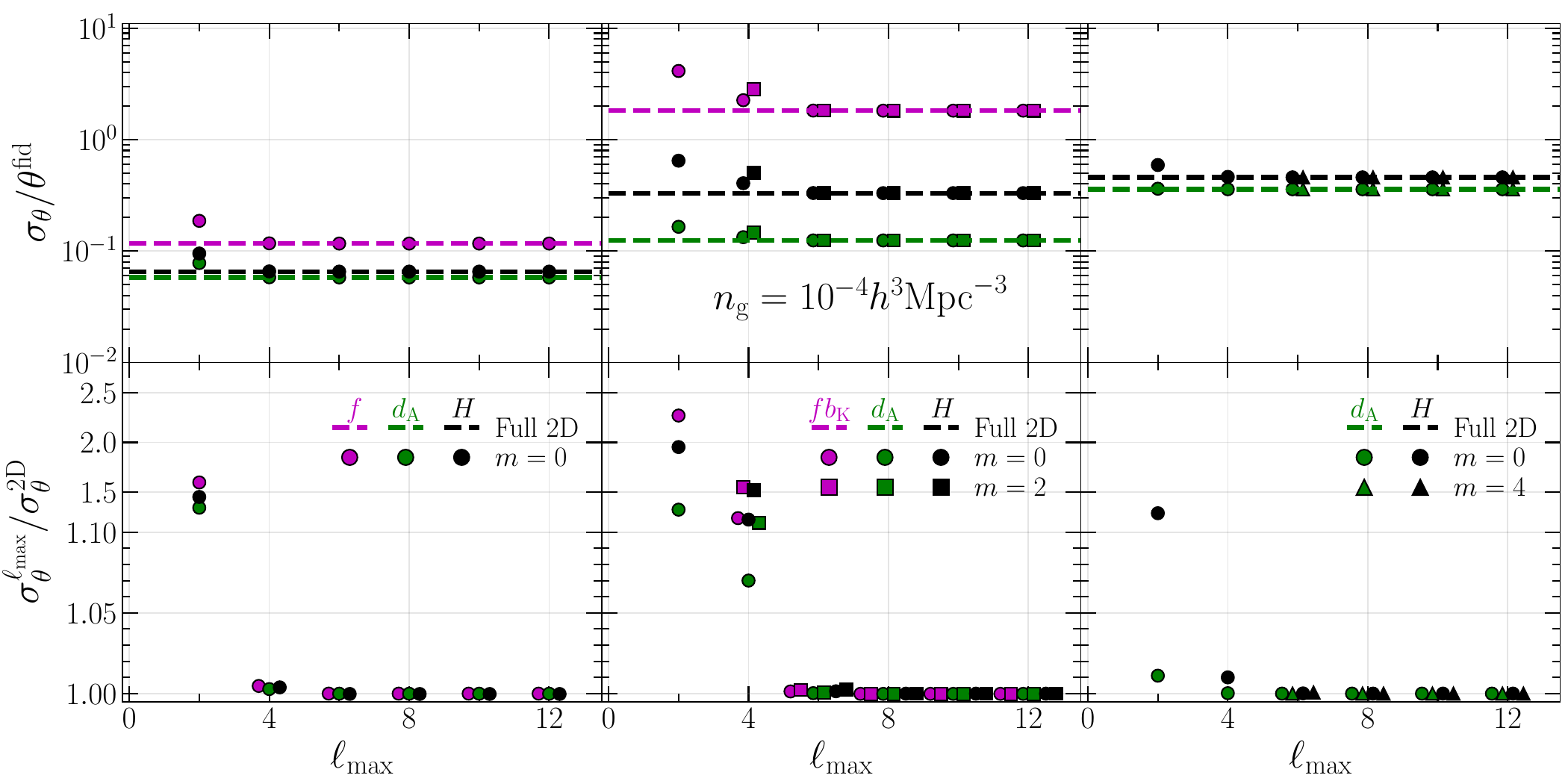}
\caption{
Same as Fig.~\ref{fig:SN} and ~\ref{fig:Fisher1D_figures} but the results expected from the low number density sample. 
Note that some symbols are slightly shifted horizontally from integer values for visibility.
}
\label{fig:lowng}
\end{figure*}
\begin{figure*}
\includegraphics[width=0.9\textwidth]{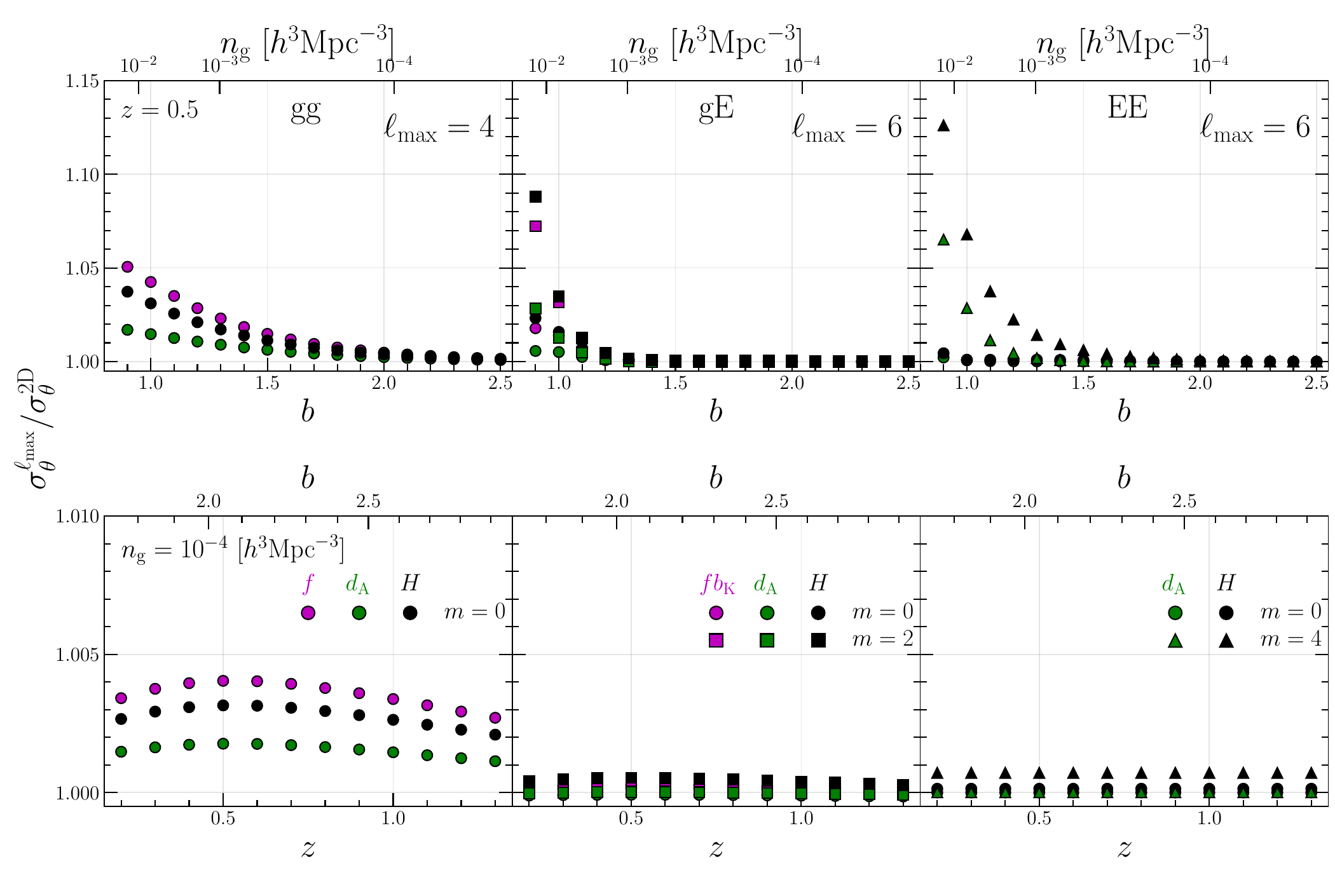}
\caption{Ratio of the one-dimensional errors on the growth rate, $f$ or $fb_{\rm K}$, and the geometric distances, $d_{\rm A}$ and $H$, obtained from the multipoles and full 2D spectra for each power spectrum, $\sigma_{\theta}^{\ell_{\rm max}}/\sigma_{\theta}^{\rm 2D}$, with $\ell_{\rm max}$ fixed at $\ell_{\rm max}=4$ for $P^{\rm gg}$ and $\ell_{\rm max}=6$ for $P^{\rm gE}$ and $P^{\rm EE}$.
The upper and lower panels show the results as a function of the bias and redshift, respectively, taking into account the number density. Here, we estimate the number density and the bias in the upper x-axis of each panel using the Sheth-Tormen mass function."}
\label{fig:Fisher_b_z_l6}
\end{figure*}
\begin{figure*}[t]
\centering
\includegraphics[width=0.49\textwidth]{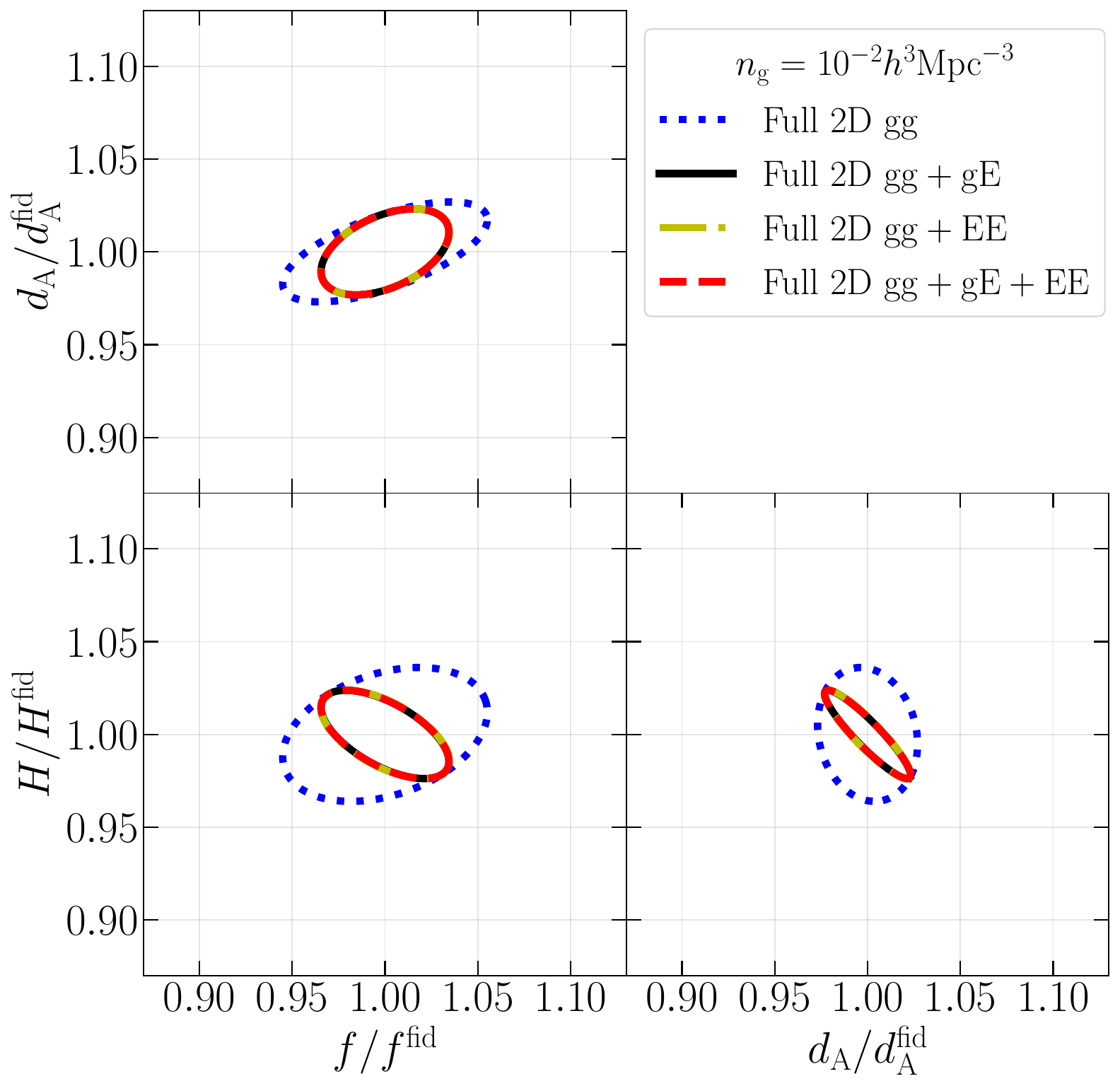}
\hfill
\includegraphics[width=0.49\textwidth]{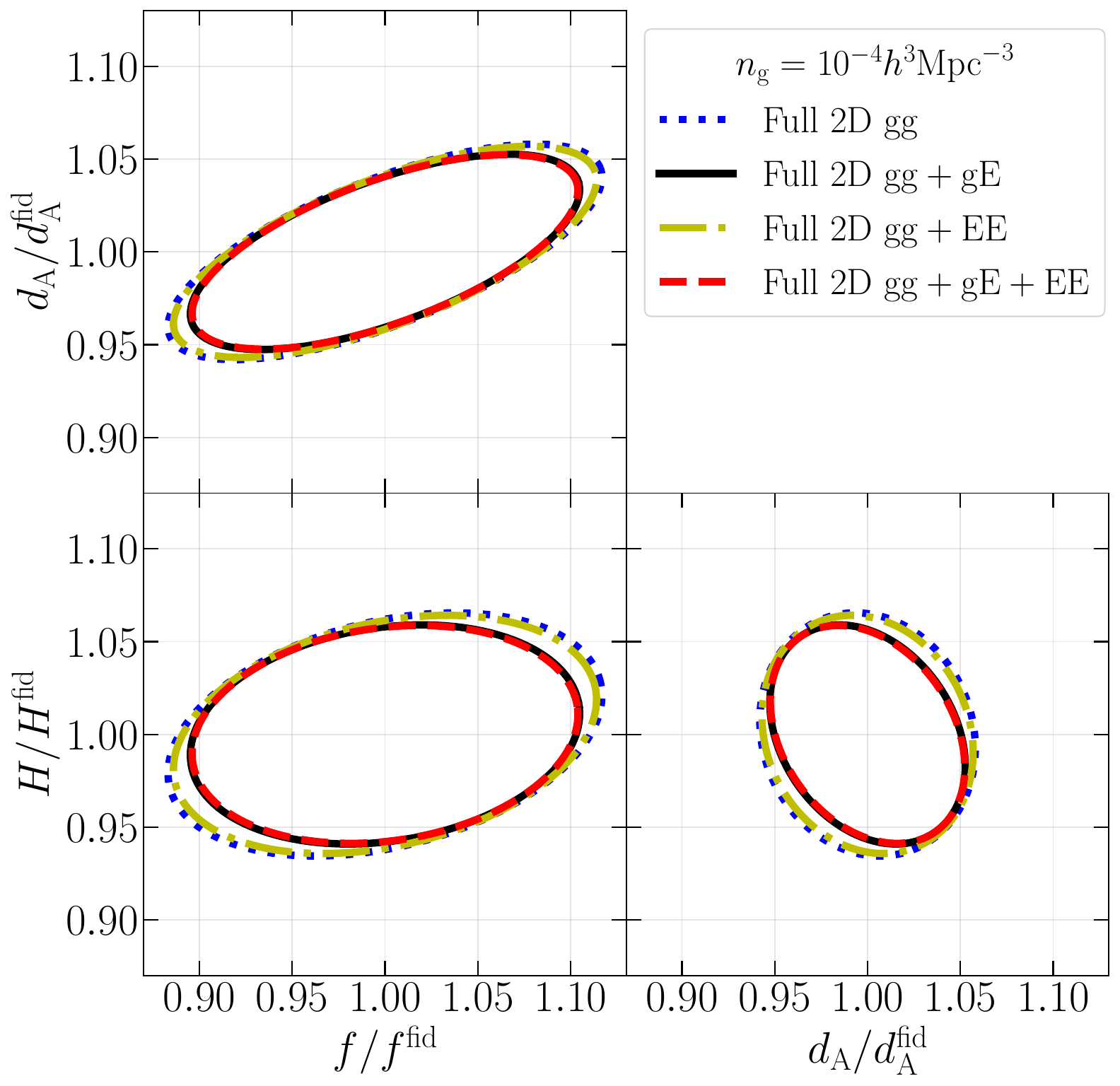}
\caption{{\it Left set}: two-dimensional marginalized errors (68 \% C.L.) on the growth rate $f$, the angular diameter distance $d_{\rm A}$ and Hubble parameter $H$, expected from the high number density sample. The blue contours indicate the results obtained from the full 2D spectra of $P^{\rm gg}$ alone, while the black, yellow, and red contours from its combinations with the IA power spectra, $P^{\rm gg}+P^{\rm gE}$, $P^{\rm gg}+P^{\rm EE}$, and $P^{\rm gg}+P^{\rm gE}+P^{\rm EE}$, respectively. {\it Right set}: Same as the left panels but the results expected from the low number density sample. 
}
\label{fig:  2D_error_2D_gg_gE_EE}
\end{figure*}

\subsection{Parameter constraints}
\label{sec: parameter constraints}

As mentioned in the previous section, we have information in vanishing higher-order multipoles via the off-diagonal components of the covariance matrix, and therefore it should affect the parameter constraints as well. 
In this subsection, we estimate the errors on the parameters obtained from the multipoles of $P^{\rm gg}$ in the Legendre basis and the IA power spectra, $P^{\rm gE}$ and $P^{\rm EE}$, in the two different bases. Here we consider the anisotropies induced not only from the RSD and projection of the galaxy shapes, but also from the AP effect described in Sec.~\ref{sec:APeffect}. 

For the Fisher matrix calculation, free parameters considered here are $\bm{\theta}=(b,f,d_{\rm A}/d^{\rm  fid}_{\rm A}, H/H^{\rm fid})$ for $P^{\rm gg}$, $\bm{\theta}=(bb_{\rm K},fb_{\rm K}, d_{\rm A}/d^{\rm fid}_{\rm A}, H/H^{\rm fid})$ for $P^{\rm gE}$, and $\bm{\theta}=(b_{\rm K}, d_{\rm A}/d^{\rm fid}_{\rm A}, H/H^{\rm fid})$ for $P^{\rm EE}$.
Note that with $P^{\rm gE}$ alone,  we cannot determine $f$ because it degenerates with $b_{\rm K}$. We are interested in the parameters that carry cosmological information, i.e., $f$, $fb_{\rm K}$, $H$, and $d_{\rm A}$. Hence, we below marginalize over the others as nuisance parameters, and present the results for those four parameters.

The top panels of Fig.~\ref{fig:  2D_error} show the two-dimensional marginalized error ellipses (68 \% C.L.) on the geometric distances ($d_{\rm A}$, $H$) forecasted from the multipoles up to $\ell=6$ in the Legendre basis and full 2D spectra of $P^{\rm gg}$, $P^{\rm gE}$, and $P^{\rm EE}$. 
Although the higher-order multipoles $\ell \geq 4$ are noisy, if measured accurately, they have  non-negligible impacts on the parameter constraints thanks to their different degeneracy directions from the other lower-order multipoles as discussed in the following.
Obviously, a single multipole moment in each power spectrum alone cannot determine $d_{\rm A}$ and $H$ due to the degeneracy between them. However, combining the $\ell=0$ and $2$ moments, namely choosing $\ell_{\rm max}=2$, can provide the simultaneous constraints on the geometric distances and adding further the $\ell=4$ moment, $\ell_{\rm max}=4$, yields a tighter constraint as clearly shown in the results of $P^{\rm gg}$ and $P^{\rm gE}$ \citep{2008PhRvD..77l3540P, 2011PhRvD..83j3527T}.

Next, the bottom panels of Fig.~\ref{fig:  2D_error} show the results in the associated Legendre basis. 
When comparing the results of the same-order multipoles between the two bases, the degeneracy directions are different. This is because the multipole moment in the associated Legendre basis can be expressed by the several multipole moments in the Legendre basis, as shown in Appendix \ref{app: mathematical relations}.
An interesting aspect in the results is the contribution from the additional non-vanishing multipole moment, namely $\ell=6$, caused by the AP effect. This multipole plays an important role in obtaining the tightest constraints when the associated Legendre basis is adopted.
Particularly, $\widetilde{P}^{\rm EE}_{4,4}$ alone cannot determine $d_{\rm A}$ and $H$, while $\widetilde{P}^{\rm EE}_{6,4}$ can help to break the degeneracy between those parameters.
Another important finding is that the $\ell=6$ moment of the IA power spectra gives tighter geometrical constraints than the $\ell=4$ moment. 

We further investigate how well multipole moments of each power spectrum, $P^{\rm gg}$, $P^{\rm gE}$ and $P^{\rm EE}$, can constrain the parameters as a function of $\ell_{\rm max}$.
The top panels in Fig.~\ref{fig:Fisher1D_figures} show the one-dimensional fractional errors on $f$, $fb_{\rm K}$, $d_{\rm A}$, and $H$. 
The bottom panels show the ratio of the errors from the multipole and full 2D spectra, $\sigma_{\theta}^{\ell_{\rm max}}/\sigma_{\theta}^{\rm 2D}$. We do not show the results obtained only from the lowest-order multipole contribution in all the bases because the degeneracy between $d_{\rm A}$ and $H$ cannot be broken as shown in Fig.\ref{fig:  2D_error}.

For $P^{\rm gg}$, the use of multipoles up to $\ell=4$ or $6$ can provide the geometrical and dynamical constraints comparable to the full 2D case, in good agreement with Ref.~\cite{2011PhRvD..83j3527T}. 
Similarly to Fig.~\ref{fig:SN}, the results of the multipole spectra for the high number density sample in the associated Legendre basis converge more slowly to the full 2D information than those in the conventional Legendre one. 
Notably, the one-dimensional errors on $H$ obtained from $P^{\rm gE}$ and $P^{\rm EE}$ are respectively about $6 \%$ and $10 \%$ larger than the full 2D case even at $\ell_{\rm max}=6$, although the derivatives of the signal vanish at $\ell>6$.

\begin{figure*}[t]
\centering
\includegraphics[width=0.9\textwidth]{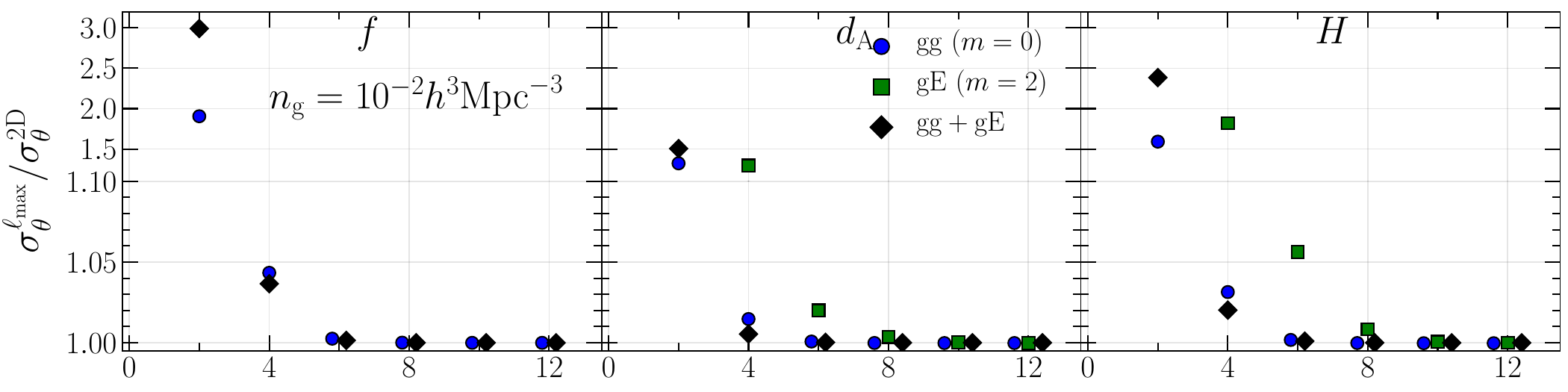}
\vfill
\includegraphics[width=0.9\textwidth]{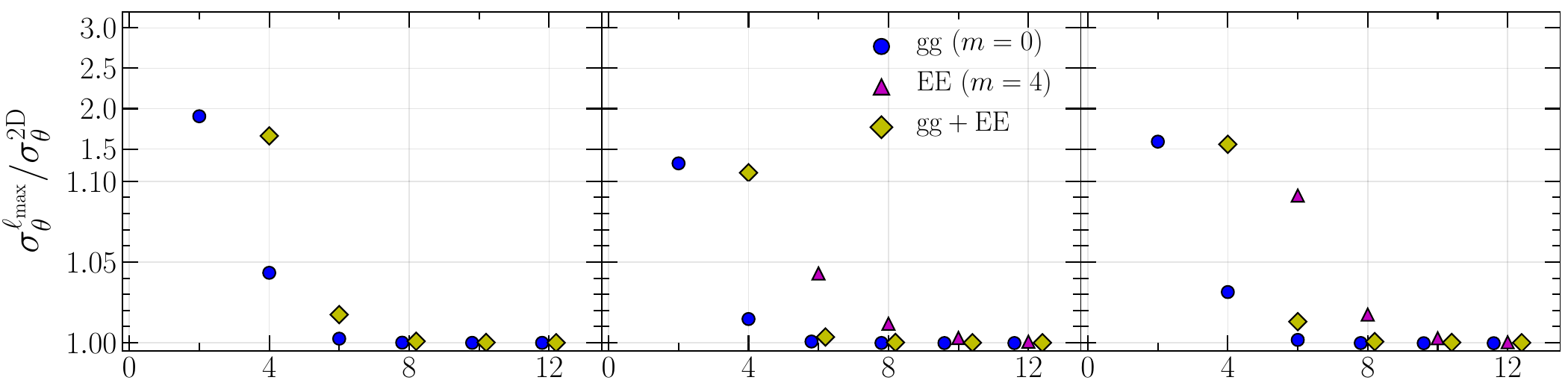}
\vfill
\includegraphics[width=0.9\textwidth]{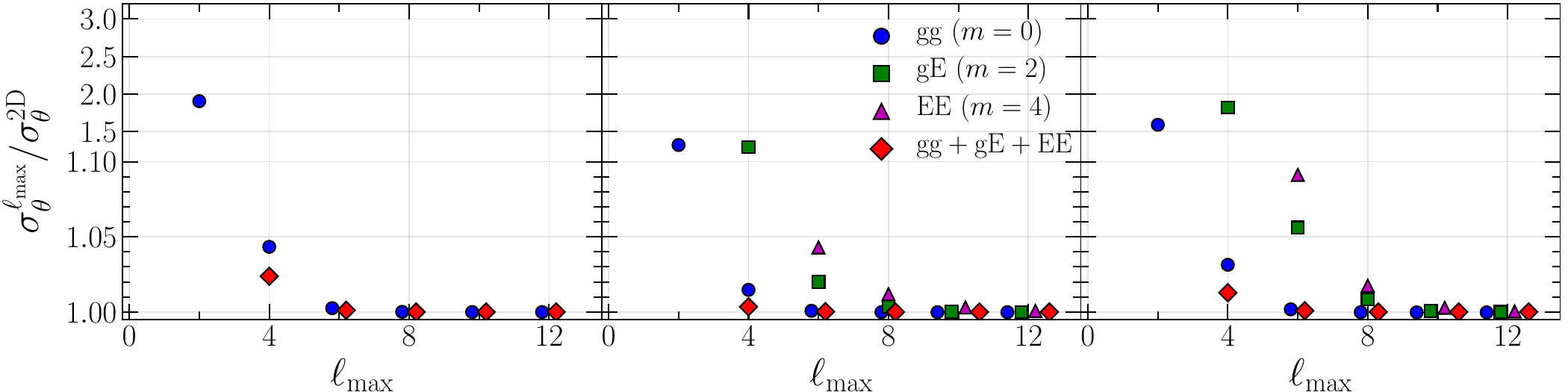}
\caption{Ratios of the one-dimensional marginalized errors on the growth rate $f$ (left), the angular diameter distance $d_{\rm A}$ (middle), and the Hubble parameter $H$ (right) obtained from the multipoles and full 2D spectra. From the top to bottom panels, we show the results for the combinations of different statistics, $P^{\rm gg} + P^{\rm gE}$, $P^{\rm gg} + P^{\rm EE}$, and $P^{\rm gg} + P^{\rm gE} + P^{\rm EE}$. For reference, we show the results from the individual statistics by the circles, squares, and triangles. Note that the vertical axis mixes different linear scalings.
}
\label{fig:Fisher1D_figures_combinations}
\end{figure*}

\section{Discussion}
\label{sec: discussion}
\subsection{Number density dependence}
\label{sec: low_ng}
\subsubsection{High vs low number density}
\label{sec: high_vs_low_ng}

In Sec.~\ref{sec: results} we discussed the results expected from the high number density sample. 
Here we present the same analysis for the low number density sample and compare the constraining power as a function of $\ell_{\rm max}$.
Fig.~\ref{fig:lowng} shows the results of the same analysis as in Figs.~\ref{fig:SN} and \ref{fig:Fisher1D_figures} but for the low number density.
In this case the multipoles of the IA power spectra up to $\ell = 4$ contain contributions comparable to the full 2D case. This is because the observed IA spectra are more affected by the shot noise than by the cosmic variance for the low number density sample.
Similar to the signal-to-noise ratios, the constraints obtained from the multipoles of the IA power spectra quickly converge to the full 2D case.   
\subsubsection{Impact of bias and redshift}
\label{sec: high_vs_low_ng}
We investigate the convergence by taking into account the changes in bias and redshift, along with the corresponding changes in number density. In Fig.~\ref{fig:Fisher_b_z_l6}, we plot the ratio of the one-dimensional errors on each parameter, obtained from the multipoles and full 2D spectra, with $\ell_{\rm max}$ fixed at $\ell_{\rm max}=4$ for $P^{\rm gg}$ and $\ell_{\rm max}=6$ for $P^{\rm gE}$ and $P^{\rm EE}$. The upper and lower panels respectively show the bias and redshift dependence on the results. 
Here, using the Sheth $\&$ Tormen mass function \cite{Sheth_1999}, $n_{\rm g}$ in the upper row is estimated from the bias and redshift, while $b$ in the lower row is from the number density and redshift. 
For the upper panel, the redshift is fixed to $z=0.5$. We find that the convergence slows down when considering the sample with high number density or low bias. For the lower panel, the number density is fixed to $n_{\rm g}=10^{-4}[h^3 {\rm Mpc}^{-3}]$. 
In this case, the impact of the bias and redshift on the convergence, not as significant as that of the number density. The changes in number density do not affect the trend of the results.
Therefore, we confirm that the convergence of the constraining power to the full 2D case mainly depends on the number density.

\subsection{Combinations of the different statistics}
\label{sec: 2D_gg_gE_EE}

Here we present the parameter constraints from combinations of the different statistics. We here examine three combinations: $P^{\rm gg} + P^{\rm gE}$, $P^{\rm gg} + P^{\rm EE}$, and $P^{\rm gg} + P^{\rm gE} + P^{\rm EE}$. To jointly analyze multiple statistics, we fully take into account the cross-covariance part (Eq.~(\ref{eq: def_all_cov})). We compute the Fisher matrix, taking the five parameters $\bm{\theta}=(b,f, b_{\rm K}, d_{\rm A}/d^{\rm fid}_{\rm A},H/H^{\rm fid})$ to be free.
Note that the covariance matrix of the multiple statistics becomes ill-defined if we consider only the CV $\times$ CV term. 
It gives the vanishing determinant because the different statistics have the same information and are completely correlated. 

Fig.~\ref{fig: 2D_error_2D_gg_gE_EE} presents the two-dimensional marginalized errors on $f$, $d_{\rm A}$ and $H$.
No matter how we combine the galaxy clustering statistics with IA statistics, the constraints are improved compared to the galaxy clustering alone, since the IA statistics tightly constrain the geometric distances, $d_{\rm A}$ and $H$, thereby breaking the degeneracy between the geometric distances and $f$ \cite{2020ApJ...891L..42T}.
The improvement for the high number density sample is evident regardless of the choice for the combinations of IA statistics. On the other hand, for the low number density sample, the observed IA auto-power spectrum, $P^{\rm EE}$, is more severely affected by the shot noise, resulting in little improvement for combination of $P^{\rm gg}+P^{\rm EE}$, as shown in the yellow dash-dotted contours of the right panels.

Finally, considering the combinations of the multiple statistics, we investigate how fast the information content of the finite multipoles converges to the full 2D case. Fig.~\ref{fig:Fisher1D_figures_combinations} shows the ratio of the one-dimensional errors on $f$, $d_{\rm A}$, and $H$ between the multipole and full 2D spectra, $\sigma_{\theta}^{\ell_{\rm max}}/\sigma_{\theta}^{\rm 2D}$, obtained by the individual statistics as well as the three cases of the combinations. 
We find that the use of multipoles up to $\ell=4$ or $6$ gives the information comparable to the full 2D case even when we adopt the associated Legendre basis to expand the IA correlations. 

\section{Conclusions}
\label{conclusion: conclusion}

In statistical data analysis of galaxy redshift surveys, one often invokes the multipole expansion in order to characterize the anisotropies along the line-of-sight direction. 
In this paper, based on the forecast studies, signal-to-noise ratios and Fisher matrix, we have shown the information content of multipole moments for the IA power spectra, considering the two different bases in the multipole expansions.

In our analysis, we have used the multipole moments of the IA power spectra and their Gaussian covariance matrices in the linear theory limit, expanded by the normalized associated Legend polynomials.
As a representative example, we considered a sample with the high number density of $10^{-2}[h^3{\rm Mpc}^{-3}]$ at $z=0.5$. 
In this work, we have newly found that the multipole contributions of the IA power spectra in the associated Legendre basis ($m>0$) slowly converge to the full 2D cases. 
It occurs because even if the higher-order multipole vanishes, its covariance does not and the vanishing multipole would contribute to the non-vanishing ones at lower order. 
We have shown that the errors on the Hubble parameter, $H$, obtained from multipoles of the IA cross- and auto-power spectra expanded in terms of the associated Legendre basis are about $6 \%$ and $10 \%$ larger than the full 2D case even when using multipoles up to $\ell=6$, although the multipole with $\ell=6$ is zero. The convergence was found faster for the lower-number density sample.
Our results show that the choice of the basis for multipole expansions is important for precise parameter constraints, since it changes the information encoded in the cosmological observables. 

Finally we comment on the modelling and assumptions used in this study. We adopted the linear theory for the power spectra and their analytic covariance matrices.
Beyond the linear theory, i.e., when considering the non-linear effects, the multipole moments of the IA spectra are non-zero even at $\ell>4$ \cite{2011PhRvD..83j3527T, 2004PhRvD..70h3007S, 2024PhRvD.109j3501O}. Thus, we will further investigate how the non-linear effects change the information contained in the multipoles of the IA spectra in our future work.
In this work, we have also assumed the plane-parallel approximation, which caused the contributions from the higher-order multipoles in the covariance.  
As pointed out by Ref.~\cite{2020MNRAS.498L..77S} (see also Ref.~\cite{Shiraishi_2021}), one can avoid these contributions by adopting the Tripolar spherical harmonics (TripoSH) as a basis in the multipole expansions.
Hence, it is necessary to investigate the information content encoded in the multipoles of the IA power spectra using a more general basis such as the TripoSH and we leave its detail investigation to our future work.

\begin{acknowledgments}
TO acknowledges support from the Ministry of Science and Technology of Taiwan under Grants No. MOST 111-2112-M-001-061- and No. NSTC 112-2112-M-001-034- and the Career Development Award, Academia Sinica (AS-CDA-108-M02) for the period of 2019-2023. This work is supported by the Japan Society for the Promotion of Science (JSPS) KAKENHI Grant No. JP23K19050 and No. 24K17043 (S. S.) and in part by MEXT/JSPS KAKENHI Grants No. JP20H05861 and No. JP21H01081 (A. T.).
\end{acknowledgments}

\appendix
\section{Useful relations}
\label{app: mathematical relations}

In this appendix, we present the derivation of useful relations regarding the multipole moments.

First, we derive the relations between the IA power spectrum and its multipoles, expanded in terms of the normalized associated Legendre basis.
The expansion coefficients, $\widetilde P^{\rm g E}_{\ell,m}$ and $\widetilde P^{\rm EE}_{\ell,m}$, satisfy the following relations
\begin{align}
\sum^{4}_{\ell=0}\sqrt{\frac{2\ell+1}{2}} \widetilde P^{\rm g E}_{\ell,0}(k)&=\sum^{4}_{\ell=0}\sqrt{\frac{2\ell+1}{2}}\widetilde P^{\rm E E}_{\ell,0}(k)=0, \label{eq:sumPl} 
\end{align}
\begin{align}
\int^{1}_{-1}d\mu \left\{P^{\rm g E}(k,\mu)\right\}^2&=\sum^{4}_{\ell=0}\left\{\widetilde P^{\rm g E}_{\ell,0}(k)\right\}^2 \nn \\
&=\sum^{4}_{\ell=2}\left\{\widetilde P^{\rm g E}_{\ell,2}(k)\right\}^2
, \label{eq:sumPGElm}
\\
\int^{1}_{-1}d\mu \left\{P^{\rm E E}(k,\mu)\right\}^2&=\sum^{4}_{\ell=0}\left\{\widetilde P^{\rm E E}_{\ell,0}(k)\right\}^2\nn \\ 
&=\left\{\widetilde P^{\rm E E}_{4,4}(k)\right\}^2
. \label{eq:sumPEElm}
\end{align}
Similar relations also hold for the correlation function.
These relations simplify the ratio of the signal-to-noise ratios obtained from the multipoles at a given $\ell_{\rm max}$ and the full 2D spectra, discussed in Sec.~\ref{sec: detectability}. We note that Ref.~\cite{2023MNRAS.518.4976S} derived Eq.~(\ref{eq:sumPl}) for the multipoles of the galaxy density-ellipticity cross-correlation function in terms of the conventional Legendre basis $\mathcal{L}_{\ell}(\mu)$ (see Appendix A in their paper). The derivation of Eqs.~(\ref{eq:sumPGElm}) and ~(\ref{eq:sumPEElm}), known as the Parseval's identity, is given below (see e.g., Ref.~\cite{1988qtam.book.....V} for the spherical harmonics case). 

Let us start by expanding an arbitrary function $f(\mu)$ by the normalized associated Legendre polynomials $\Theta_{\ell,m}(\mu)$,
\begin{equation}
f(\mu)=\sum_{\ell\geq m}f_{\ell,m}\Theta_{\ell,m}(\mu)
, \label{eq:any function}
\end{equation}
where $\ell$ and $m$ are integers with $\ell \geq |m|$ and the latter is arbitrarily chosen.
In our case, this function $f$ is either $\widetilde P^{\rm g E}_{\ell,m}$ or $\widetilde P^{\rm EE}_{\ell,m}$.
Squaring both sides of this equation yields
\begin{equation}
\left\{f(\mu)\right\}^2=\sum_{\ell\geq m}\sum_{\ell'\geq m} f_{\ell,m}f_{\ell',m}\Theta_{\ell,m}(\mu)\Theta_{\ell',m}(\mu)
. \label{eq:any function square}
\end{equation}
Integrating both sides from -1 to 1 over $\mu$, we obtain
\begin{align}
\int^{1}_{-1}d\mu\left\{f(\mu)\right\}^2&=\sum_{\ell\geq m}\sum_{\ell'\geq m} f_{\ell,m}f_{\ell',m} \nn \\ 
&\qquad \times \int^{1}_{-1}d\mu\Theta_{\ell,m}(\mu)\Theta_{\ell',m}(\mu)\notag \\
&=\sum_{\ell\geq m}\sum_{\ell'\geq m}f_{\ell,m}f_{\ell',m}\delta^{\rm K}_{\ell,\ell'}\notag \\
&=\sum_{\ell\geq m}\left(f_{\ell,m}\right)^2
, \label{eq:int}
\end{align}
where we used the orthonormal condition of $\Theta_{\ell,m}$. Obviously, this equation holds for arbitrary $m$.

Next, we derive the relation of the multipole coefficients between the two bases. 
The multipole moment in the associated Legendre basis is calculated through 
\begin{align}
f_{\ell, m} &=\int^{1}_{-1}d\mu f(\mu) \Theta_{\ell,m}(\mu)
, \label{eq:multipole moment}
\end{align}
Substituting Eq~\ref{eq:any function}) into Eq~\ref{eq:multipole moment}), we obtain
\begin{align}
f_{\ell, m} 
&=\sum_{\ell'\geq m'}\int^{1}_{-1} d \mu f_{\ell',m'}\Theta_{\ell',m'}(\mu)\Theta_{\ell,m}(\mu)
\notag \\ 
&=\sum_{\ell'\geq 0}\int^{1}_{-1} d \mu f_{\ell',0}\Theta_{\ell',0}(\mu)\Theta_{\ell,m}(\mu)
, \label{eq:multipole}
\end{align}
where we take $m'=0$ corresponding to the Legendre basis in the last line. 
\\
The multipoles of the gE power spectrum in the associated Legendre basis are given by
\begin{align}
\widetilde{P}^{\rm gE}_{2, 2}&=\sqrt{\frac{5}{6}}\widetilde{P}^{\rm gE}_{0, 0}-\sqrt{\frac{1}{6}}\widetilde{P}^{\rm gE}_{2, 0},
\\
\widetilde{P}^{\rm gE}_{4, 2}&=\sqrt{\frac{1}{10}}\widetilde{P}^{\rm gE}_{0, 0}+\sqrt{\frac{1}{2}}\widetilde{P}^{\rm gE}_{2, 0}-\sqrt{\frac{2}{5}}\widetilde{P}^{\rm gE}_{4, 0}
, \label{eq:PGEl2}
\end{align}
and that of the EE power spectrum is given by
\begin{align}
\widetilde{P}^{\rm EE}_{4, 4}&=\sqrt{\frac{7}{10}}\widetilde{P}^{\rm EE}_{0, 0}-\sqrt{\frac{2}{7}}\widetilde{P}^{\rm EE}_{2, 0}+\sqrt{\frac{1}{70}}\widetilde{P}^{\rm EE}_{4, 0}
. \label{eq:PEEl4}
\end{align}
The multipole coefficient in the associated Legendre basis is expressed by the several coefficients in the Legendre basis (see Ref.~\cite{2024PhRvD.109j3501O} for the correlation function case taking into account the non-linear effects).
\begin{widetext}
\section{Multipole covariance}
\label{app: derivation cov}
In Sec.~\ref{sec:covariance}, we provided the formula of the multipole covariance in terms of the normalized associated Legendre polynomials $\Theta_{\ell,m}$. 
In this appendix, we show the explicit expressions of the auto- and cross-covariance matrices for the $\rm gE$ and $\rm EE$ power spectrum multipoles. 

By integrating Eq.~(\ref{eq:CovPlm}) over $\mu$ using Eq.~(\ref{eq:Gaunt}), we obtain the multipole coefficients satisfying the selection rules of the Wigner 3-j symbols and orthonormality condition of $\Theta_{\ell,m}$, which are given as
\begin{align}
{\rm COV}_{\ell 2, \ell'2}^{\rm g E, g E}(k)
&=\frac{2}{N_k} \Biggl[ \frac{32}{15015} \sqrt{\frac{7(2 \ell + 1) (2 \ell' + 1)}{2}}\Biggl\{3 \sqrt{5}\left(143 b^2 + 26 b f + 3 f^2\right) 
\left(
\begin{array}{ccc}
\ell & \ell'& 4\\
2 & 2 & -4
\end{array}
\right)
\left(
\begin{array}{ccc}
\ell & \ell'& 4\\
0 & 0 & 0
\end{array}
\right) \notag \\ &
+ 52 f\left(5b + f\right) 
\left(
\begin{array}{ccc}
\ell & \ell'& 6\\
2 & 2 & -4
\end{array}
\right)
\left(
\begin{array}{ccc}
\ell & \ell'& 6\\
0 & 0 & 0
\end{array}
\right)
+ 8 \sqrt{11} f^2
\left(
\begin{array}{ccc}
\ell & \ell'& 8\\
2 & 2 & -4
\end{array} 
\right)
\left(
\begin{array}{ccc}
\ell & \ell'& 8\\
0 & 0 & 0
\end{array}
\right)\Biggr\}\left\{b_{\rm K}P_{\rm L}(k)\right\}^2\notag \\ & +
\Biggl\{\sigma^2_{\gamma}\left(b^2+\frac{2}{3}bf+\frac{1}{5}f^2\right)\delta^{\rm K}_{\ell,\ell'}+\frac{4}{105}\sqrt{\left(2\ell+1\right)\left(2\ell'+1\right)}  \sum_{L}\sqrt{2L+1}
\left(
\begin{array}{ccc}
\ell & \ell'& L\\
2 & 2 & -4
\end{array}
\right)
\left(
\begin{array}{ccc}
\ell & \ell'& L\\
0 & 0 & 0
\end{array}
\right)
\notag \\ &
\times \Bigl\{f \sigma_{\gamma}^2 \Bigl(\sqrt{5}\left(7b+3f\right)\mathcal{I}^{L,-4}_{2, 0}+2f\mathcal{I}^{L,-4}_{4, 0}\Bigr)  +
2 \sqrt{70} b_{\rm K}^2 \mathcal{I}^{L,-4}_{4, 4}\Bigr\}
\Biggr\}\frac{P_{\rm L}(k)}{n_{\rm g}}
+
\frac{\sigma_{\gamma}^2}{n_{\rm g}^2}\delta^{\rm K}_{{\ell},{\ell'}}
\Biggr] , \label{eq:COVgE}
\end{align}
for the auto-covariance between the $\rm gE$ power spectrum multipoles with $m=2$, 
\begin{align}
{\rm COV}_{\ell 2, \ell'4}^{\rm g E, E E}(k)
&=\frac{2}{N_k} \Biggl[ \frac{64}{15} \sqrt{\left(2 \ell + 1\right) \left(2 \ell' + 1\right)} \Biggl\{\frac{1}{\sqrt{231}}\left(15 b  +  f\right) 
\left(
\begin{array}{ccc}
\ell & \ell'& 6\\
2 & 4 & -6
\end{array}
\right)
\left(
\begin{array}{ccc}
\ell & \ell'& 6\\
0 & 0 & 0
\end{array}
\right)
+ \frac{2}{\sqrt{429}} f
\left(
\begin{array}{ccc}
\ell & \ell'& 8\\
2 & 4 & -6
\end{array}
\right) \notag \\ & \times
\left(
\begin{array}{ccc}
\ell & \ell'& 8\\
0 & 0 & 0
\end{array}
\right)\Biggr\}b_{{\rm K}}^3\left\{P_{\rm L}(k)\right\}^2
 +
\Biggl\{\frac{8}{7}\sqrt{\frac{\left(2\ell+1\right)\left(2\ell'+1\right)}{30}}\sum_{L}\sqrt{2L+1}
\left(
\begin{array}{ccc}
\ell & \ell'& L\\
2 & 4 & -6
\end{array}
\right)
\left(
\begin{array}{ccc}
\ell & \ell'& L\\
0 & 0 & 0
\end{array}
\right)
\notag \\ & \times
\Bigl\{\left(7b+f\right) \mathcal{I}^{L,-6}_{2, 2}+\frac{2 \sqrt{3}}{3}f\mathcal{I}^{L,-6}_{4, 2}\Bigr\}\Biggr\} 
b_{\rm K}P_{\rm L}(k)\frac{\sigma_{\gamma}^2}{n_{\rm g}}
\Biggr] , \label{eq:COVgEEE}
\end{align}
for the cross-covariance between the $\rm gE$ and $\rm EE$ power spectrum multipoles with $m=2$ and $4$, respectively, and
\begin{align}
{\rm COV}_{\ell 4, \ell'4}^{\rm E E, E E}(k)
&=\frac{4}{N_k}\Biggl[\frac{256}{3} \sqrt{\frac{(2 \ell + 1) (2 \ell' + 1)}{1430}}
\left(
\begin{array}{ccc}
\ell & \ell'& 8\\
4 & 4 & -8
\end{array}
\right)
\left(
\begin{array}{ccc}
\ell & \ell'& 8\\
0 & 0 & 0
\end{array}
\right)\left\{b_{{\rm K}}^2P_{\rm L}(k)\right\}^2 +
\frac{32}{3} \sqrt{\frac{\left(2 \ell + 1 \right) \left(2 \ell' + 1 \right)}{70}}\notag \\ & \times \sum_{L} \sqrt{2L+1}
\left(
\begin{array}{ccc}
\ell & \ell'& L\\
4 & 4 & -8
\end{array}
\right) 
\left(
\begin{array}{ccc}
\ell & \ell'& L\\
0 & 0 & 0
\end{array}
\right)\mathcal{I}^{L,-8}_{4, 4}
b_{{\rm K}}^2P_{\rm L}(k)\frac{\sigma_{\gamma}^2}{n_{\rm g}}
+
\frac{\sigma_{\gamma}^4}{n_{\rm g}^2}\delta^{\rm K}_{{\ell},{\ell'}}
\Biggr] , \label{eq:COVEE}
\end{align}
for the auto-covariance between the $\rm EE$ spectrum multipoles with $m=4$.
In the above expressions, we define the quantity $\mathcal{I}^{L,M}_{\ell, m}$ as follows:
\begin{align}
\mathcal{I}^{L,M}_{\ell, m}=\int_{-1}^1 d\mu \Theta_{L,M}(\mu) \Theta_{\ell,m}(\mu), \label{eq:int theta}
\end{align}
where the condition $M=m$ leads to the orthonormality relation, $\mathcal{I}^{L,m}_{\ell, m} = \delta^{\rm K}_{L,\ell}$.

For reference, we provide the multipole covariance matrices up to $\ell=4$ in terms of the associated Legendre basis as follows:
\begin{align}
{\rm COV}_{2 2, 2 2}^{\rm gE, gE}(k)
&= 
\frac{2}{3003 N_k}\Biggl[32  \left(143 b^2+26 b f+3 f^2\right)\left\{b_{\rm K}P_{\rm L}(k)\right\}^2
+ 
143\Biggl\{\frac{16}{{n_{\rm g}}}b_{\rm K}^2+\left(21 b^2+6 b f+f^2\right) \frac{\sigma^2_{\rm \gamma}}{n_{\rm g}}\Biggr\}P_{\rm L}(k) \notag \\ &
\qquad \qquad \qquad + 
\frac{3003\sigma^2_{\rm \gamma}}{n^2_{\rm g}}\Biggr]
 ,
\label{eq: COVP2222}
\\
{\rm COV}_{22,42}^{\rm g E, g E}(k)
&= 
\frac{8}{1001 \sqrt{3} N_k}\Biggl[-16\left(13 b^2-4 b f-f^2\right)\left\{b_{\rm K}P_{\rm L}(k)\right\}^2
+ 13 \Biggl\{-\frac{8}{n_{\rm g}} b_{\rm K}^2+f \left(11 b+3 f\right)\frac{\sigma^2_{\rm \gamma}}{n_{\rm g}}\Biggr\}P_{\rm L}(k)\Biggr]
,
\label{eq: COVP2242}
\\
{\rm COV}_{42,42}^{\rm gE, gE}(k)
&= 
\frac{2}{17017 N_k}\Biggl[
32 \left(459 b^2+170 b f+39 f^2\right)\left\{b_{\rm K}P_{\rm L}(k)\right\}^2
+ 
17\Biggl\{\frac{432}{n_{\rm g}}
b_{{\rm K}}^2+\left(1001 b^2+806 b f+237 f^2\right) \frac{\sigma^2_{\rm \gamma}}{n_{\rm g}}\Biggr\}\notag \\ &
\qquad \qquad \qquad
\times P_{\rm L}(k)
+ 
\frac{17017\sigma^2_{\rm \gamma}}{n^2_{\rm g}}\Biggr]
,
\label{eq: COVP4242}
\end{align}
for the auto-covariance of the $\rm gE$ power spectra, 
\begin{align}
{\rm COV}_{22, 44}^{\rm g E, E E}(k)
&= 
\frac{16}{143\sqrt{21}N_k}\Biggl[
8 \left(15b+f\right) b_{\rm K}^3 \left\{P_{\rm L}(k)\right\}^2
+ 
13 \left(11b+f\right)b_{\rm K}P_{\rm L}(k)\frac{\sigma^2_{\rm \gamma}}{n_{\rm g}}
\Biggr]
,
\label{eq: COVGIIIP2244}
\\
{\rm COV}_{42, 44}^{\rm g E, E E}(k)
&= 
\frac{32}{2431\sqrt{7}N_k}\Biggl[
8 \left(-34b+f\right) b_{\rm K}^3 \left\{P_{\rm L}(k)\right\}^2
+ 
17 \left(-13b+2f\right)b_{\rm K}P_{\rm L}(k)\frac{\sigma^2_{\rm \gamma}}{n_{\rm g}}
\Biggr]
,
\label{eq: COVGIIIP4244}
\end{align}
for the cross-covariance of the $\rm gE$ and $\rm EE$ power spectra. 
\begin{align}
{\rm COV}_{44, 44}^{\rm E E, E E}(k)
&= 
\frac{4}{2431N_k}\Biggl[
1792 \left\{b_{\rm K}^2 P_{\rm L}(k)\right\}^2
+ 
4080 b_{\rm K}^2P_{\rm L}(k)\frac{\sigma^2_{\rm \gamma}}{n_{\rm g}}
+ 
2431\left(\frac{\sigma^2_{\rm \gamma}}{n_{\rm g}}\right)^2\Biggr]
,
\label{eq: COVIIP4444}
\end{align}
for the auto-covariance of the $\rm EE$ power spectra. 
As we mentioned in Sec.~\ref{sec:covariance}, the multipole coefficients of the covariances generally appear even at higher-order moments where the power spectrum multipoles themselves vanish.
\end{widetext}

\section{Derivatives of power spectra}
\label{app: dev_power_spectra}
In this appendix, we present the derivatives of the full 2D power spectra and power spectrum multipoles with respect to their free parameters, which we used in the Fisher analysis. 
The observed power spectra $P^{\rm XY}(k,\mu)$ (${\rm X,Y={g \ or \ E}}$) distorted by the linear RSD and AP effects are generally expressed as $P^{\rm XY, obs}(k, \mu)=(\alpha_{\parallel}/\alpha^2_{\perp})P^{\rm XY}(q,\nu)$,
where $\alpha_{\perp} \equiv d_{\rm A}/d^{\rm fid}_{\rm A}$ and $\alpha_{\parallel} \equiv H/H^{\rm fid}$.
The superscripts ``obs'' and ``fid'' respectively denote observed and fiducial quantities. The new variables $q$ and $\nu$ are respectively defined by
$q \equiv \sqrt{q^2_{\parallel}+q^2_{\perp}}=k\sqrt{\alpha^{-2}_{\perp}+(\alpha^2_{\parallel}-\alpha^{-2}_{\perp})\mu^2}$
and
$\nu \equiv q_{\parallel}/q=\alpha_{\parallel} \mu /\sqrt{\alpha^{-2}_{\perp}+(\alpha^2_{\parallel}-\alpha^{-2}_{\perp})\mu^2},$
where $k_{\perp}$ and $k_{\parallel}$ are, respectively, related to the true ones ($q_{\perp}$, $q_{\parallel}$) by $k_{\perp}=k\sqrt{1-\mu^2}=\alpha_{\perp} q_{\perp}$ and $k_{\parallel}=k \mu= q_{\parallel}/\alpha_{\parallel}$.

By differentiating each power spectrum with respect to their free parameters, namely $\partial P^{\rm gg, obs}/\partial \theta$ with $\theta=(b, f, \alpha_{\perp}, \alpha_{\parallel})$, $\partial P^{\rm gE, obs}/\partial \theta$ with $\theta=(bb_{\rm K}, fb_{\rm K}, \alpha_{\perp}, \alpha_{\parallel})$, and $\partial P^{\rm EE, obs}/\partial \theta$ with $\theta=(b_{\rm K}, \alpha_{\perp}, \alpha_{\parallel})$, at the fiducial values $(b^{\rm fid}, f^{\rm fid}, b_{\rm K}^{\rm fid}, \alpha^{\rm fid}_{\perp}, \alpha^{\rm fid}_{\parallel})$ , we have 
\begin{align}
\left.\frac{\partial P^{\rm gg, obs}}{\partial b}\right|_{\rm fid}&= 2\left(b+f \mu^2\right)P_{\rm L}(k), \label{eq: dPgg_b} \\
\left.\frac{\partial P^{\rm gg, obs}}{\partial f}\right|_{\rm fid}&= 2 \mu^2 \left(b+f \mu^2\right)P_{\rm L}(k), \label{eq: dPgg_f} \\
\left.\frac{\partial P^{\rm gg, obs}}{\partial \alpha_{\perp}}\right|_{\rm fid} &= \left(b+f \mu ^2\right)\Bigl[-b\left(\dlnPk+2\right)+\left\{b\dlnPk+f\left(2-\dlnPk\right)\right\} \notag \\&\times \mu^2 + f\left(\dlnPk-4\right)\mu^4 \Bigr]P_{\rm L}(k), \label{eq: dPgg_dA} \\
\left.\frac{\partial P^{\rm gg,  obs}}{\partial \alpha_{\parallel}}\right|_{\rm fid}&= \left(b+f \mu ^2\right) \left\{b+\left(5f+b \dlnPk \right)\mu^2 \right. \notag \\ & + \left. f \left(\dlnPk-4\right)\mu^4\right\}P_{\rm L}(k), \label{eq: dPgg_H} 
\end{align}
for the derivatives of the $\rm gg$ auto-power spectrum,
\begin{align}
\left.\frac{\partial P^{\rm gE, obs}}{\partial bb_{\rm K}}\right|_{\rm fid}&= \left(1-\mu^2\right)P_{\rm L}(k), \label{eq: dPgE_bbk}  \\
\left.\frac{\partial P^{\rm gE, obs}}{\partial fb_{\rm K}}\right|_{\rm fid}&= \mu^2\left(1-\mu^2\right)P_{\rm L}(k), \label{eq: dPgE_fbk} \\
\left.\frac{\partial P^{\rm gE, obs}}{\partial \alpha_{\perp}}\right|_{\rm fid}&= b_{\rm K}\left(1-\mu ^2\right)\Bigl[-b\left(\dlnPk+2\right)\notag \\&+\left\{b\left(\dlnPk-2\right)-f\dlnPk\right\}\mu^2
+f\left(\dlnPk-4\right)\mu^4\Bigr]
P_{\rm L}(k), \label{eq: dPgE_dA} \\
\left.\frac{\partial P^{\rm gE, obs}}{\partial \alpha_{\parallel}}\right|_{\rm fid}&=b_{\rm K}\left(1-\mu ^2\right)\Bigl[b+\left\{b\left(\dlnPk-2\right)+3f\right\}\mu^2 \notag \\ &
+f\left(\dlnPk-4\right)\mu^4\Bigr]
P_{\rm L}(k), \label{eq: dPgE_H} 
\end{align}
for the derivatives of the $\rm gE$ cross-power spectrum, and
\begin{align}
\left.\frac{\partial P^{\rm EE, obs}}{\partial b_{\rm K}}\right|_{\rm fid}&= 2 b_{\rm K} \left(1-\mu^2\right)^2P_{\rm L}(k), \label{eq: dPEE_bk}  \\
\left.\frac{\partial P^{\rm EE, obs}}{\partial \alpha_{\perp}}\right|_{\rm fid}&= b_{\rm K}^2 \left(1-\mu ^2\right)^2\left\{-\dlnPk-2+ \left(\dlnPk-4\right)\mu^2\right\}P_{\rm L}(k), \label{eq: dPEE_dA} \\
\left.\frac{\partial P^{\rm EE, obs}}{\partial \alpha_{\parallel}}\right|_{\rm fid}&= b_{\rm K}^2 \left(1-\mu ^2\right)^2\left\{1+ \left(\dlnPk-4\right)\mu^2\right\}P_{\rm L}(k), \label{eq: dPEE_H}
\end{align}
for the derivatives of the $\rm EE$ auto-power spectrum.
Here, we define the logarithmic derivatives of $P_{\rm L}(k)$ with respect to $k$, $n \equiv \partial \ln P_{\rm L}(k)/\partial \ln k$. In the above expressions, we already substituted $\alpha^{\rm fid}_{\perp}=\alpha^{ \rm fid}_{\parallel}=1$ and omitted the superscript ``fid'' on the other parameters, $b$, $f$, and $b_{\rm K}$. 

With the above expressions, we compute the derivatives of the multipole moments with respect to the free parameters in terms of the normalized associated Legendre polynomials $\Theta_{\ell,m}$ by
\begin{align}
\left.\frac{\partial \widetilde{P}^{\rm XY}_{\ell,m}}{\partial \theta}\right|_{\rm fid} &=\int^{1}_{-1}{d}\mu \Theta_{\ell,m}(\mu) \left. \frac{\partial P^{\rm XY}}{\partial \theta}\right|_{\rm fid}
. \label{eq: dplm def}
\end{align}
In the following, we present the derivatives of the non-vanishing gg, gE, and EE power spectrum multipoles for $m=0$, $m=0,2$, and $m=0,4$, respectively (see Sec.~\ref{sec:multipole power spectrum} for the corresponding multipoles).

First, we show the derivatives of the non-vanishing $\rm gg$~auto-power spectrum multipoles with respect to the parameters $(b, f, \alpha_{\perp}, \alpha_{\parallel})$ for $m=0$:
\begin{align}
&
\left(\left.\frac{\partial \widetilde{P}^{\rm gg, obs}_{0,0}}{\partial b}\right|_{\rm fid}, \left.\frac{\partial \widetilde{P}^{\rm gg, obs}_{2,0}}{\partial b}\right|_{\rm fid}\right) \notag \\ &
=\left( \frac{2}{3} \sqrt{2} \left(3 b+f\right) P_{\rm L}(k) ,\,
\frac{4}{3} \sqrt{\frac{2}{5}} f P_{\rm L}(k)\right) , \label{eq: dpgglm defb}
\end{align}
\begin{align}
&
\left(\left.\frac{\partial \widetilde{P}^{\rm gg, obs}_{0,0}}{\partial f}\right|_{\rm fid}, \left.\frac{\partial \widetilde{P}^{\rm gg, obs}_{2,0}}{\partial f}\right|_{\rm fid}, \left.\frac{\partial \widetilde{P}^{\rm gg, obs}_{4,0}}{\partial f}\right|_{\rm fid}\right) \notag \\ &=
\left(\frac{2}{15} \sqrt{2} \left(5 b+3 f\right) P_{\rm L}(k), \right.  \frac{4}{21} \sqrt{\frac{2}{5}} \left(7 b+6 f\right) P_{\rm L}(k), \notag \\& \qquad \left.\frac{16}{105} \sqrt{2} f P_{\rm L}(k)\right), \label{eq: dpgglm deff}
\end{align}
\begin{align}
&
\left(\left.\frac{\partial \widetilde{P}^{\rm gg, obs}_{0,0}}{\partial \alpha_{\perp}}\right|_{\rm fid}, \left.\frac{\partial \widetilde{P}^{\rm gg, obs}_{2,0}}{\partial \alpha_{\perp}}\right|_{\rm fid}, \left.\frac{\partial \widetilde{P}^{\rm gg, obs}_{4,0}}{\partial \alpha_{\perp}}\right|_{\rm fid},\left.\frac{\partial \widetilde{P}^{\rm gg, obs}_{6,0}}{\partial \alpha_{\perp}}\right|_{\rm fid}\right) \notag \\ &=
\left(-\frac{2 \sqrt{2}}{105}\left(\dlnPk+3\right)\left(35 b^2+14 b f+3 f^2\right) P_{\rm L}(k), \right. \notag \\ & \qquad \frac{2}{21} \sqrt{\frac{2}{5}}\left\{7b^2\dlnPk-2bf\left(\dlnPk+12\right)-f^2\left(\dlnPk+8\right)\right\}P_{\rm L}(k), \notag \\& \qquad  \left.\frac{16 \sqrt{2} f}{1155}\left\{11b\left(\dlnPk-2\right) +f\left(2\dlnPk-19\right)\right\}P_{\rm L}(k), \right. \notag \\ & \qquad \left.\frac{16}{231} \sqrt{\frac{2}{13}} f^2 \left(\dlnPk-4\right)P_{\rm L}(k) \right), \label{eq: dpgglm defdA}
\end{align}
\begin{align}
&
\left(\left.\frac{\partial \widetilde{P}^{\rm gg, obs}_{0,0}}{\partial \alpha_{\parallel}}\right|_{\rm fid}, \left.\frac{\partial \widetilde{P}^{\rm gg, obs}_{2,0}}{\partial \alpha_{\parallel}}\right|_{\rm fid}, \left.\frac{\partial \widetilde{P}^{\rm gg, obs}_{4,0}}{\partial \alpha_{\parallel}}\right|_{\rm fid},\left.\frac{\partial \widetilde{P}^{\rm gg, obs}_{6,0}}{\partial \alpha_{\parallel}}\right|_{\rm fid}\right) \notag \\ &=
\left(\frac{\sqrt{2}}{105}\left(\dlnPk+3\right)\left(35 b^2+42 b f+15 f^2\right) P_{\rm L}(k), \right. \notag \\ & \qquad \frac{2}{21} \sqrt{\frac{2}{5}}\left\{7b^2\dlnPk+6bf\left(2\dlnPk+3\right)+5f^2\left(\dlnPk+2\right)\right\}P_{\rm L}(k), \notag \\& \qquad \left.\frac{8 \sqrt{2} f}{1155}\left\{22b\left(\dlnPk-2\right) +5f\left(3\dlnPk-1\right)\right\}P_{\rm L}(k), \right. \notag \\ & \qquad \left.\frac{16}{231} \sqrt{\frac{2}{13}} f^2 \left(\dlnPk-4\right)P_{\rm L}(k) \right). \label{eq: dpgglm defH}
\end{align}
Next, we show the derivatives of non-vanishing $\rm gE$ cross-power spectrum multipoles with respect to the parameters $(bb_{\rm K}, fb_{\rm K}, \alpha_{\perp}, \alpha_{\parallel})$:
\begin{align}
&
\left(\left.\frac{\partial \widetilde{P}^{\rm gE, obs}_{0,0}}{\partial bb_{\rm K}}\right|_{\rm fid}, \left.\frac{\partial \widetilde{P}^{\rm gE, obs}_{2,0}}{\partial bb_{\rm K}}\right|_{\rm fid}\right)
\notag \\
&=\left( \frac{2}{3} \sqrt{2} P_{\rm L}(k), -\frac{2}{3} \sqrt{\frac{2}{5}}P_{\rm L}(k)\right), \label{eq: dpgElm defbbk}
\end{align}
\begin{align}
&
\left(\left.\frac{\partial \widetilde{P}^{\rm gE, obs}_{0,0}}{\partial fb_{\rm K}}\right|_{\rm fid}, \left.\frac{\partial \widetilde{P}^{\rm gE, obs}_{2,0}}{\partial fb_{\rm K}}\right|_{\rm fid},\left.\frac{\partial \widetilde{P}^{\rm gE, obs}_{4,0}}{\partial fb_{\rm K}}\right|_{\rm fid}\right)
\notag \\ &=\left( \frac{2}{15} \sqrt{2} P_{\rm L}(k), \frac{2}{21} \sqrt{\frac{2}{5}} P_{\rm L}(k), -\frac{8}{105} \sqrt{2} P_{\rm L}(k)\right), \label{eq: dpgElm deffbk}
\end{align}
\begin{align}
&
\left(\left.\frac{\partial \widetilde{P}^{\rm gE, obs}_{0,0}}{\partial \alpha_{\perp}}\right|_{\rm fid}, \left.\frac{\partial \widetilde{P}^{\rm gE, obs}_{2,0}}{\partial \alpha_{\perp}}\right|_{\rm fid}, \left.\frac{\partial \widetilde{P}^{\rm gE, obs}_{4,0}}{\partial \alpha_{\perp}}\right|_{\rm fid},\left.\frac{\partial \widetilde{P}^{\rm gE, obs}_{6,0}}{\partial \alpha_{\perp}}\right|_{\rm fid}\right) \notag \\ &=
\left(-\frac{8\sqrt{2}}{105}\left(\dlnPk+3\right) (7b+f)b_{\rm K}  P_{\rm L}(k), \right. \notag \\ & \qquad \frac{8}{21} \sqrt{\frac{2}{5}}\left\{b\left(2\dlnPk+3\right)-f\right\}b_{\rm K}P_{\rm L}(k), \notag \\& \qquad \left.\frac{8 \sqrt{2}}{1155}\left\{-11b\left(\dlnPk-2\right)+f\left(7\dlnPk+16\right)\right\} b_{\rm K}P_{\rm L}(k), \right. \notag \\ & \qquad \left.-\frac{16}{231} \sqrt{\frac{2}{13}} f \left(\dlnPk-4\right)b_{\rm K}P_{\rm L}(k) \right), \label{eq: dpgEl0 defdA}
\end{align}
\begin{align}
&
\left(\left.\frac{\partial \widetilde{P}^{\rm gE, obs}_{0,0}}{\partial \alpha_{\parallel}}\right|_{\rm fid}, \left.\frac{\partial \widetilde{P}^{\rm gE, obs}_{2,0}}{\partial \alpha_{\parallel}}\right|_{\rm fid}, \left.\frac{\partial \widetilde{P}^{\rm gE, obs}_{4,0}}{\partial \alpha_{\parallel}}\right|_{\rm fid},\left.\frac{\partial \widetilde{P}^{\rm gE, obs}_{6,0}}{\partial \alpha_{\parallel}}\right|_{\rm fid}\right) \notag \\ &=
\left(\frac{2\sqrt{2}}{105}\left(\dlnPk+3\right) (7b+3f)b_{\rm K}  P_{\rm L}(k), \right. \notag \\ & \qquad \frac{2}{21} \sqrt{\frac{2}{5}}\left\{b\left(\dlnPk-9\right)+f\left(\dlnPk-1\right)\right\}b_{\rm K}P_{\rm L}(k), \notag \\& \qquad \left.-\frac{8 \sqrt{2}}{1155}\left\{11b\left(\dlnPk-2\right)+f\left(4\dlnPk+17\right)\right\} b_{\rm K}P_{\rm L}(k), \right. \notag \\ & \qquad \left.-\frac{16}{231} \sqrt{\frac{2}{13}} f \left(\dlnPk-4\right)b_{\rm K}P_{\rm L}(k) \right), \label{eq: dpgEl0 defH}
\end{align}
for $m=0$, and
\begin{align}
\left.\frac{\partial \widetilde{P}^{\rm gE, obs}_{2,2}}{\partial bb_{\rm K}}\right|_{\rm fid}=\frac{4 P_{\rm L}(k)}{\sqrt{15}}, \label{eq: dpgEl2 defbbk}
\end{align}
\begin{align}
&
\left(\left.\frac{\partial \widetilde{P}^{\rm gE, obs}_{2,2}}{\partial fb_{\rm K}}\right|_{\rm fid}, \left.\frac{\partial \widetilde{P}^{\rm gE, obs}_{4,2}}{\partial fb_{\rm K}}\right|_{\rm fid}\right)
=\left( \frac{4 P_{\rm L}(k)}{7 \sqrt{15}}, \frac{8 P_{\rm L}(k)}{21 \sqrt{5}}\right), \label{eq: dpgEl2 deffbk}
\end{align}
\begin{align}
&
\left(\left.\frac{\partial \widetilde{P}^{\rm gE, obs}_{2,2}}{\partial \alpha_{\perp}}\right|_{\rm fid}, \left.\frac{\partial \widetilde{P}^{\rm gE, obs}_{4,2}}{\partial \alpha_{\perp}}\right|_{\rm fid}, \left.\frac{\partial \widetilde{P}^{\rm gE, obs}_{6,2}}{\partial \alpha_{\perp}}\right|_{\rm fid}\right) \notag \\ &=
\left(-\frac{8}{21 \sqrt{15}}\left\{3b\left(3\dlnPk+8\right)+f\left(\dlnPk+2\right)\right\}b_{\rm K}P_{\rm L}(k), \right. \notag \\ & \qquad \frac{8}{231 \sqrt{5}} \left\{11b\left(\dlnPk-2\right)-f\left(5\dlnPk+24\right)\right\} b_{\rm K}P_{\rm L}(k), \notag \\& \qquad \left.\frac{32}{33} \sqrt{\frac{2}{1365}} f \left(\dlnPk-4\right)b_{\rm K} P_{\rm L}(k)\right), \label{eq: dpgEl2 defdA}
\end{align}
\begin{align}
&
\left(\left.\frac{\partial \widetilde{P}^{\rm gE, obs}_{2,2}}{\partial \alpha_{\parallel}}\right|_{\rm fid}, \left.\frac{\partial \widetilde{P}^{\rm gE, obs}_{4,2}}{\partial \alpha_{\parallel}}\right|_{\rm fid}, \left.\frac{\partial \widetilde{P}^{\rm gE, obs}_{6,2}}{\partial \alpha_{\parallel}}\right|_{\rm fid}\right) \notag \\ &=
\left(\frac{4}{21 \sqrt{15}}\left(3 b+f\right)\left(\dlnPk+5\right) b_{\rm K}P_{\rm L}(k), \right. \notag \\ & \qquad \frac{8}{231 \sqrt{5}} \left\{11b\left(\dlnPk-2\right)+3f\left(2\dlnPk+3\right)\right\} b_{\rm K}P_{\rm L}(k), \notag \\& \qquad \left.\frac{32}{33} \sqrt{\frac{2}{1365}} f \left(\dlnPk-4\right)b_{\rm K} P_{\rm L}(k)\right), \label{eq: dpgEl2 defH}
\end{align}
for $m=2$.
Finally, we show the derivatives of non-vanishing $\rm EE$ auto-power spectrum multipoles with respect to the parameters $(b_{\rm K}, \alpha_{\perp}, \alpha_{\parallel})$:
\begin{align}
&
\left(\left.\frac{\partial \widetilde{P}^{\rm EE, obs}_{0,0}}{\partial b_{\rm K}}\right|_{\rm fid}, \left.\frac{\partial \widetilde{P}^{\rm EE, obs}_{2,0}}{\partial b_{\rm K}}\right|_{\rm fid},\left.\frac{\partial \widetilde{P}^{\rm EE, obs}_{4,0}}{\partial b_{\rm K}}\right|_{\rm fid}\right)
\notag \\ &=\left(\frac{16}{15} \sqrt{2} b_{\rm K} P_{\rm L}(k), -\frac{32}{21} \sqrt{\frac{2}{5}} b_{\rm K} P_{\rm L}(k), \frac{16}{105} \sqrt{2} b_{\rm K} P_{\rm L}(k)\right), \label{eq: dpEEl0 defbk}
\end{align}
\begin{align}
&
\left(\left.\frac{\partial \widetilde{P}^{\rm EE, obs}_{0,0}}{\partial \alpha_{\perp}}\right|_{\rm fid}, \left.\frac{\partial \widetilde{P}^{\rm EE, obs}_{2,0}}{\partial \alpha_{\perp}}\right|_{\rm fid}, \left.\frac{\partial \widetilde{P}^{\rm EE, obs}_{4,0}}{\partial \alpha_{\perp}}\right|_{\rm fid},\left.\frac{\partial \widetilde{P}^{\rm EE, obs}_{6,0}}{\partial \alpha_{\perp}}\right|_{\rm fid}\right) \notag \\ &=
\left(-\frac{16\sqrt{2}}{35}\left(\dlnPk+3\right)b_{\rm K}^2P_{\rm L}(k), \right. \frac{16}{21}\sqrt{\frac{2}{5}} \left(\dlnPk+2\right)b_{\rm K}^2 P_{\rm L}(k), \notag \\& \quad \left.-\frac{16\sqrt{2}}{385}\left(3\dlnPk-1\right) b_{\rm K}^2P_{\rm L}(k), \right. \left.\frac{16}{231} \sqrt{\frac{2}{13}}\left(\dlnPk-4\right)b_{\rm K}^2P_{\rm L}(k)\right), \label{eq: dpEEl0 defdA}
\end{align}
\begin{align}
&
\left(\left.\frac{\partial \widetilde{P}^{\rm EE, obs}_{0,0}}{\partial \alpha_{\parallel}}\right|_{\rm fid}, \left.\frac{\partial \widetilde{P}^{\rm EE, obs}_{2,0}}{\partial \alpha_{\parallel}}\right|_{\rm fid}, \left.\frac{\partial \widetilde{P}^{\rm EE, obs}_{4,0}}{\partial \alpha_{\parallel}}\right|_{\rm fid},\left.\frac{\partial \widetilde{P}^{\rm EE, obs}_{6,0}}{\partial \alpha_{\parallel}}\right|_{\rm fid}\right) \notag \\ &=
\left(\frac{8\sqrt{2}}{105}\left(\dlnPk+3\right)b_{\rm K}^2P_{\rm L}(k), \right. -\frac{16}{21} \sqrt{\frac{2}{5}} b_{\rm K}^2 P_{\rm L}(k), \notag \\& \quad \left.-\frac{8\sqrt{2}}{1155} \left(7\dlnPk-39\right)b_{\rm K}^2P_{\rm L}(k), \right. \left.\frac{16}{231} \sqrt{\frac{2}{13}}\left(\dlnPk-4\right)b_{\rm K}^2P_{\rm L}(k)\right), \label{eq: dpEEl0 defH}
\end{align}
for $m=0$, and
\begin{align}
\left.\frac{\partial \widetilde{P}^{\rm EE, obs}_{4,4}}{\partial b_{\rm K}}\right|_{\rm fid}=\frac{32}{3 \sqrt{35}} b_{\rm K} P_{\rm L}(k), \label{eq: dpEEl4 defbk}
\end{align}
\begin{align}
&
\left(\left.\frac{\partial \widetilde{P}^{\rm EE, obs}_{4,4}}{\partial \alpha_{\perp}}\right|_{\rm fid}, \left.\frac{\partial \widetilde{P}^{\rm EE, obs}_{6,4}}{\partial \alpha_{\perp}}\right|_{\rm fid}\right) \notag \\ &=
\left(-\frac{32}{33\sqrt{35}}\left(5\dlnPk+13\right) b_{\rm K}^2 P_{\rm L}(k), \right. \notag \\& \qquad \left.\frac{32}{33 \sqrt{91}}\left( \dlnPk -4\right) b_{\rm K}^2P_{\rm L}(k)\right), \label{eq: dpEEl4 defdA}
\end{align}
\begin{align}
&
\left(\left.\frac{\partial \widetilde{P}^{\rm EE, obs}_{4,4}}{\partial \alpha_{\parallel}}\right|_{\rm fid}, \left.\frac{\partial \widetilde{P}^{\rm EE, obs}_{6,4}}{\partial \alpha_{\parallel}}\right|_{\rm fid}\right) \notag \\ &=
\left(\frac{16}{33 \sqrt{35}}\left(\dlnPk+7\right)b_{\rm K}^2P_{\rm L}(k), \right. \notag \\& \qquad \left. \frac{32}{33 \sqrt{91}}\left( \dlnPk -4\right) b_{\rm K}^2P_{\rm L}(k)\right), \label{eq: dpEEl4 defH}
\end{align}
for $m=4$. 
We note that, due to the AP effect, the power spectrum multipoles and their derivatives with respect to the parameters are non-zero for $\ell>6$ even in the linear theory limit. However, after substituting the fiducial values into their expressions, the multipoles $\ell > 4$ for the power spectrum and $\ell > 6$ for their derivatives vanish.

\bibliography{basis}


\end{document}